\documentclass[11pt]{article}

\usepackage[T1,T2A]{fontenc}
\usepackage[cp1251]{inputenc}
\usepackage{textcomp}
\usepackage[centertags]{amsmath}
\usepackage{amsfonts}
\usepackage{amssymb}
\usepackage{revsymb}
\usepackage[pdftex]{hyperref}
\usepackage{graphicx}
\usepackage{graphbox}
\usepackage[numbers,sort&compress]{natbib}

\usepackage{paperinitial}

\paperinitialization{15mm}{15mm}{15mm}{15mm}{2pt}{10pt}

\DeclareMathOperator{\Ai}{Ai}

\DeclareMathOperator{\Sp}{Sp}

\newcommand{\lan}{\langle}
\newcommand{\ran}{\rangle}

\newcommand{\e}{\varepsilon}
\newcommand{\vf}{\varphi}
\newcommand{\vk}{\varkappa}
\newcommand{\s}{\sigma}

\newcommand{\al}{\alpha}
\newcommand{\be}{\beta}
\newcommand{\ga}{\gamma}

\newcommand{\de}{\delta}
\newcommand{\De}{\Delta}

\newcommand{\la}{\lambda}
\newcommand{\La}{\Lambda}
\newcommand{\ups}{\upsilon}

\newcommand{\spx}{\mathbf{x}}

\newcommand{\spp}{\mathbf{p}}
\newcommand{\spk}{\mathbf{k}}
\newcommand{\spe}{\mathbf{e}}
\newcommand{\spa}{\mathbf{a}}

\begin{document}
\allowdisplaybreaks[4]
\frenchspacing
\setlength{\unitlength}{1pt}

\title{{\Large\textbf{Semiclassical probability of radiation of twisted photons in the ultrarelativistic limit}}}

\date{}

\author{O.V. Bogdanov${}^{1),2)}$\thanks{E-mail: \texttt{bov@tpu.ru}},\; P.O. Kazinski${}^{1)}$\thanks{E-mail: \texttt{kpo@phys.tsu.ru}},\; and G.Yu. Lazarenko${}^{1)}$\thanks{E-mail: \texttt{laz@phys.tsu.ru}}\\[0.5em]
{\normalsize ${}^{1)}$ Physics Faculty, Tomsk State University, Tomsk 634050, Russia}\\
{\normalsize ${}^{2)}$ Tomsk Polytechnic University, Tomsk 634050, Russia}}

\maketitle

\begin{abstract}

The semiclassical general formula for the probability of radiation of twisted photons by ultrarelativistic scalar and Dirac particles moving in the electromagnetic field of a general form is derived. This formula is the analog of the Baier-Katkov formula for the probability of radiation of one plane wave photon with the quantum recoil taken into account. The derived formula is used to describe the radiation of twisted photons by charged particles in undulators and laser waves. Thus, the general theory of undulator radiation of twisted photons and radiation of twisted photons in the nonlinear Compton process is developed with account for the quantum recoil. The explicit formulas for the probability to record a twisted photon are obtained in these cases. In particular, we found that the quantum recoil and spin degrees of freedom increase the radiation probability of twisted photons in comparison with the formula for scalar particles without recoil. In the range of applicability of the semiclassical formula, the selection rules for undulator radiation established in the purely classical framework are not violated. The manifestation of the blossoming out rose effect in the nonlinear Compton process in a strong laser wave with circular polarization and in the wiggler radiation is revealed. Several examples are studied: the radiation of MeV twisted photons by $180$ GeV electrons in the wiggler; the radiation of twisted photons by $256$ MeV electrons in strong electromagnetic waves produced by the CO$_2$ and Ti:Sa lasers; and the radiation of MeV twisted photons by $51.1$ MeV electrons in the electromagnetic wave generated by the FEL with photon energy $1$ keV.

\end{abstract}

\section{Introduction}

Nowadays the Baier-Katkov (BK) semiclassical method \cite{BaiKat1,BKMS,BaKaStr,BaKaStrbook} is a standard tool to describe radiation of plane wave photons by ultrarelativistic charged particles in external electromagnetic fields of a general form. This method effectively includes the quantum recoil experienced by a charged particle in radiating one hard photon and is applicable for the energies of radiated photons right up to the energy of the radiating particle (for other semiclassical methods see, e.g., \cite{AkhShul91,BBTq1,BBTq2,Bord.1,BBTZ,AkhShul}). The BK method is realized in several computer codes \cite{GuiBandTikh,BeKoSo14,BBGT15,WPKU18} and proved to be very successful \cite{BaKaStrbook,Wist14,DHIMT16,BSBM18,Li2018,Martins16,PiMuHaKermp.2,Poder18,Cole18}. A comparison of the radiation probability obtained by this method with the exact QED results, when they are obtainable, reveals a spectacular agreement \cite{Wist14,BaKaStrbook,BKMS,BaKaStr}. We use this method to derive the radiation probability of one twisted photon \cite{MolTerTorTor,JenSerprl,JenSerepj,Ivanov11,KnyzSerb,PadgOAM25,Roadmap16,AndBabAML,TorTorTw,AndrewsSLIA} by an ultrarelativistic charged particle with account for the quantum recoil. In the case of negligible quantum recoil, the obtained general formula reduces to the one derived in \cite{BKL2}.

According to the BK method, in the ultrarelativistic limit, the one-photon radiation probability can be calculated by means of a formula resembling the classical formula for the intensity of radiation \cite{LandLifshCTF.2,JacksonCE}. One need not solve the Dirac or Klein-Gordon equations in the given external fields but just find the solution of the Lorentz equations in these fields. The spin degrees of freedom are characterized by the spin vector and its evolution is governed by the Bargmann-Michel-Telegdi equation. If the radiation probability is summed over spin polarizations of the escaping particle and averaged over spin polarizations of the incoming one, the dynamics of the spin vector are irrelevant for evaluation of the radiation probability. All that drastically simplifies the calculations of radiation probability. Of course, there are certain limitations of this semiclassical method. The complete list of them is presented in Sec. \ref{Gener_Form}. The main idea standing behind our derivation of the radiation probability of twisted photons is to find the approximate expression for the product of radiation amplitudes using the procedure developed in \cite{BaiKat1,BKMS,BaKaStr,BaKaStrbook} and to integrate it over the azimuth angles of photon momenta with the corresponding weights. The last step creates the twisted photon in the out-state.

Having derived the general formula, we employ it to investigate the radiation probability of twisted photons by electrons in undulators (Sec. \ref{Undul}) and strong laser pulses (Sec. \ref{Laser}). Currently, the twisted photons of different spectral ranges are used in fundamental science and technology \cite{KnyzSerb,PadgOAM25,Roadmap16,AndBabAML,TorTorTw,AndrewsSLIA}. In the optical range and below, the detectors are designed allowing to decompose an arbitrary electromagnetic radiation into twisted photons \cite{LPBFAC,BLCBP,SSDFGCY,LavCourPad,RGMMSCFR}. The formulas we obtain are aimed to describe correctly the radiation of hard twisted photons with MeV energies and above that are not accessible for a direct observation. The hard twisted photons can be employed to study the properties of nuclear matter by exciting higher multipole transitions in nuclei and hadrons (see, e.g., the discussion in \cite{JenSerprl,JenSerepj,TaHaKa,AfSeSol18}). Rather recently, it was shown that hard twisted photons can be generated in channeling radiation \cite{ABKT,EpJaZo} and strong laser pulses \cite{TaHaKa,KatohPRL,TairKato,CLHK,ZYCWS18}. In Sec. \ref{Undul}, we show that MeV photons can be produced by $180$ GeV electrons in helical wigglers. Besides, we develop a general theory of undulator radiation of twisted photons with the quantum recoil taken into account. In particular, we show that the selection rules for the forward radiation of twisted photons by helical undulators \cite{KatohPRL,TaHaKa,BKL2,SasMcNu,HMRR,HeMaRo,BHKMSS,HKDXMHR,KatohSRexp,Rubic17} are not affected by the quantum recoil, at least, in the domain of applicability of the semiclassical method. Then, in Sec. \ref{Laser}, we revisit the problem of radiation of twisted photons by electrons in laser waves studied in \cite{TaHaKa,KatohPRL,TairKato}. We generalize the results of \cite{TaHaKa,KatohPRL,TairKato} to the case of laser waves with an arbitrary amplitude envelope and include the influence of quantum recoil. Thus, we develop a general theory of twisted photon production in the nonlinear Compton process. Of course, it is just another description of the nonlinear Compton process usually formulated in terms of plane wave photons \cite{BKMS,BaKaStr,BaKaStrbook,PiMuHaKermp.2,DHIMT16,BSBM18,Cole18,Poder18,GFSh.3,SSFS,Sherwin,NikRit64,Ritus.2,IvaKotSer,GHQZAPYJ,MaPiKei,PTMK,PiHaKei2,PoKoStSt}. In Sec. \ref{Laser}, we apply this general theory and describe the radiation of twisted photons by electrons in strong laser pulses produced by the free electron laser (FEL), CO${_2}$ and Ti:Sa lasers. The obtained general formulas can be implemented in computer codes to describe the radiation of twisted photons by ultrarelativistic charged particles in electromagnetic fields of a rather general configuration, in particular, in channeling.

The paper is organized as follows. In Sec. \ref{Gener_Form}, we derive the general semiclassical formulas for the one-photon radiation probability of twisted photons by scalar and Dirac particles and discuss their applicability conditions. In Sec. \ref{Undul}, we elaborate a general theory of undulator radiation of twisted photons with account for the quantum recoil undergone by a charged particle in radiating a photon. We also obtain a simple estimate for the number of radiated twisted photons and specialize the general applicability conditions to the case of undulator radiation. As expected, the wiggler radiation of lower harmonics where the main part of twisted photons is radiated has a clear imprint of the ``blossoming out rose'' effect \cite{Bord.1}. In Sec. \ref{Laser}, we consider the radiation of twisted photons by charged particles in an intense laser wave of a circular polarization. In particular, we trace the manifestation of blossoming out rose effect in this radiation. Several examples are presented there. Some cumbersome calculations are removed to Appendices \ref{TwPlw}, \ref{Azim_Int}. In Conclusion, we summarize the results.

We use the system of units such that $\hbar=c=1$ and $e^2=4\pi\al$, where $\al$ is the fine structure constant. Besides, we vastly use the notation introduced in \cite{BKL2}.

\section{General formulas}\label{Gener_Form}

Let us begin with the case of a stationary external magnetic field. The generalization to the case of a general electromagnetic field will be given below. In the presence of the electromagnetic field, the following process is possible
\begin{equation}\label{inclus_proc}
    e^-_i\rightarrow \ga_\al+X,
\end{equation}
where $e^-_i$ is the initial electron in the state $i$, $\ga_\al$ is the twisted photon \cite{MolTerTorTor,JenSerprl,JenSerepj,Ivanov11,KnyzSerb,MHSSF,SMFSAS14,GottfYan,JaurHac,BiaBirBiaBir} recorded by the detector in the state $\al$, and $X$ denotes the rest of particles that are not recorded by the detector. The probability of such an inclusive process equals
\begin{equation}
    w(\al;i)=\sum_{X}\lan i|\hat{U}_{-\infty,\infty}|X;\ga_\al\ran\lan\ga_\al;X|\hat{U}_{\infty,-\infty}|i\ran,
\end{equation}
where $\hat{U}$ is the evolution operator of QED in the presence of the external field (see, e.g., \cite{GrMuRaf,GFSh.3}). In the first Born approximation with the exact account for the external electromagnetic field, the final state contains only one electron
\begin{equation}
    w(\al;i)\approx\sum_{f}\lan i|\hat{U}_{-\infty,\infty}|f;\ga_\al\ran\lan\ga_\al;f|\hat{U}_{\infty,-\infty}|i\ran,
\end{equation}
where $f$ characterizes the final electron state. In this approximation, using the completeness relation, we obtain
\begin{equation}
    w(\al;i)=\lan i|\hat{U}_{-\infty,\infty}|\ga_\al\ran\lan\ga_\al|\hat{U}_{\infty,-\infty}|i\ran,
\end{equation}
i.e., the probability of process \eqref{inclus_proc} is equal to the average number of photons in the final state $\al$. According to \eqref{tw_w_in_pl_w}, the twisted photon state can be decomposed into the plane wave ones \cite{MolTerTorTor,JenSerprl,JenSerepj,Ivanov11,KnyzSerb,MHSSF,SMFSAS14,GottfYan,JaurHac,BiaBirBiaBir}. Then
\begin{equation}\label{prob_gen}
    w(\al;i)=\sum_{\spk_1,\spk_2}\La^*_{\al;s,\spk_2}\La_{\al;s,\spk_1} \lan i|\hat{U}_{-\infty,\infty}|s,\spk_2\ran\lan s,\spk_1|\hat{U}_{\infty,-\infty}|i\ran,
\end{equation}
where $\La_{\al;s,\spk}$ are the coefficients of expansion \eqref{tw_w_in_pl_w}. Therefore, in the first Born approximation, $w(\al;i)$ can be found from the matrix element for the plane wave photons (see the notation in \cite{BaKaStrbook})
\begin{equation}\label{matr_elem0}
    C(\spk_2,\spk_1):=\frac{e^2}{(2\pi)^32Vk_0}\lan i| \int_{-\infty}^\infty dt_1dt_2 e^{i k_0(t_1-t_2)} \hat{M}^\dag(t_2,\spk_2) \hat{M}(t_1,\spk_1)|i\ran,
\end{equation}
which should be integrated over the azimuth angles of the vectors $\spk_{1,2}$ with the corresponding phase factors as in \eqref{tw_w_in_pl_w}. Notice that we work in the coordinate system adapted to the detector of twisted photons with the axis directed along the unit vector $\spe_3$ (see for details \cite{BKL2}). The unit vectors $\{\spe_1,\spe_2,\spe_3\}$ of this coordinate system constitute a right-hand triple. The perpendicular components of a vector are those lying in the plane spanned by $\{\spe_1,\spe_2\}$. The photons are supposed to lie on the mass shell
\begin{equation}
    k_0=\sqrt{k_{1\perp}^2+k_{13}^2}=\sqrt{k_{2\perp}^2+k_{23}^2},
\end{equation}
and
\begin{equation}\label{phot_mom}
    k_{1\perp}=k_{2\perp},\qquad k_{13}=k_{23},
\end{equation}
i.e., the photon momenta $\spk_{1,2}$ differ only by the azimuth angles. The helicities of photons with momenta $\spk_{1,2}$ are the same.

We will evaluate \eqref{matr_elem0} using the approximation introduced in \cite{BaiKat1,BKMS,BaKaStr,BaKaStrbook}. Recall that it is assumed in this approximation that
\begin{enumerate}
  \item The charged particle is ultrarelativistic, i.e., the Lorentz factor $\ga\gg1$ and $\vk/\gamma\ll1$, where $\vk=\max(1,K)$, $K=\lan\be_\perp\ran\ga$ is the undulator strength parameter, and $\lan\be_\perp\ran$ is a characteristic value of the velocity component perpendicular to the detector axis;
  \item The particle wave packet is sufficiently narrow in the momentum space;
  \item The size of the wave packet in the configuration space is small in comparison with the characteristic scale of variation of the external electromagnetic field in space;
  \item $\mathbf{n}_{1,2}:=\spk_{1,2}/k_0$ lie in the cone directed along the detector axis (the axis $3$) with the opening of order $2\vk/\gamma$, i.e., we consider the region where the main part of radiation is concentrated.
\end{enumerate}
Below, there will appear the additional restriction related to the fact that we consider the radiation of twisted photons.

If the above conditions are satisfied, then, in evaluating average \eqref{matr_elem0} in the leading order in $1/\gamma$, we can use the analog of the Thomas-Fermi approximation. Namely, the commutator
\begin{equation}\label{ultrarel_appr}
    |\lan[\hat{P}_\mu, \hat{A}_\nu]\ran|/\lan \hat{\mathbf{P}}^2\ran\sim\frac{H}{H_0\ga^2}\ll1,\qquad\ga\gg1,
\end{equation}
is small for large particle energies. Here $H$ is the characteristic magnitude of the electromagnetic field, $H_0$ is the critical (Schwinger) field, and $\hat{P}_\mu=\hat{p}_\mu-e A_\mu(\hat{\spx})$. Hence, the noncommutativity of operators $\hat{\spx}$ and $\hat{\mathbf{P}}$ entering into the operators $\hat{M}^\dag(t_2,\spk_2)$, $\hat{M}(t_1,\spk_1)$ can be neglected. At the same time, the noncommutativity of $\hat{\mathbf{P}}$ and $e^{-i\spk\hat{\spx}}$ in $\hat{M}$ cannot be ignored because the exponent is a rapidly varying function for $|\spk|\sim\e$, where $\e$ is the energy of a charged particle.

\paragraph{Scalar particle.}

For a scalar charged particle, we have \cite{BaKaStrbook}
\begin{equation}\label{M_matr}
    \hat{M}(t_1,\spk_1)=\hat{P}_0^{-1/2}(\mathbf{f}^*_1\hat{\mathbf{P}}(t_1))e^{-i\spk_1\hat{\spx}(t_1)}\hat{P}_0^{-1/2},\qquad \hat{M}^\dag(t_2,\spk_2)=\hat{P}_0^{-1/2}(\mathbf{f}_2\hat{\mathbf{P}}(t_2))e^{i\spk_2\hat{\spx}(t_2)}\hat{P}_0^{-1/2},
\end{equation}
where $\mathbf{f}_{1,2}=\mathbf{f}(\spk_{1,2})$ are the polarization vectors of physical photons. All the operators in \eqref{M_matr} are given in the Heisenberg representation with the Hamiltonian
\begin{equation}\label{qHam}
    \hat{H}=\sqrt{\hat{\mathbf{P}}^2+m^2}+eA_0(\hat{\spx}),
\end{equation}
Notice that $A_0=0$ in the case of time-independent magnetic field. However, we leave $A_0$ intact since its presence influences the derivation only in one point (Eq. \eqref{2exp1}), which we shall discuss separately below. In order to avoid additional complications with the vacuum definition (see, e.g., \cite{GrMuRaf,GFSh.3,MigdalMM,Migdal}), we suppose that
\begin{equation}\label{subcr_field}
    |eA_0|\lesssim m,
\end{equation}
which is fulfilled for all the electromagnetic fields achievable at the present moment in laboratories. Therefore, $P_0=m\gamma\approx\e$ up to the terms of order $1/\gamma$.

Let us carry the exponent entering into $\hat{M}^\dag(t_2,\spk_2)$ to the right and the one entering into $\hat{M}(t_1,\spk_1)$ to the left:
\begin{equation}
    \hat{M}^\dag(t_2,\spk_2) \hat{M}(t_1,\spk_1)=\hat{P}_{02}^{-1/2}(\mathbf{f}_2 \hat{\mathbf{P}}_2)\hat{P}_{02}'^{-1/2} e^{i\spk_2 \hat{\spx}_2} e^{-i\spk_1 \hat{\spx}_1} \hat{P}_{01}'^{-1/2}(\mathbf{f}^*_1 \hat{\mathbf{P}}_1) \hat{P}_{01}^{-1/2},
\end{equation}
where
\begin{equation}
    \hat{P}'_{0\,1,2}=P_0(\hat{\mathbf{P}}'_{1,2}),\qquad \hat{\mathbf{P}}'_{1,2}:=\hat{\mathbf{P}}_{1,2}-\spk_{1,2},
\end{equation}
and $\hat{\mathbf{P}}_{1,2}\equiv\hat{\mathbf{P}}(t_{1,2})$. Then we transform the operator expression
\begin{equation}\label{2exp}
    e^{i\spk_2 \hat{\spx}_2}e^{-i\spk_1 \hat{\spx}_1}= e^{i\spk_{\perp 2} \hat{\spx}_{\perp 2}} e^{i k_3 \hat{x}_{32}} e^{-i k_3 \hat{x}_{31}} e^{-i\spk_{\perp1} \hat{\spx}_{\perp1}}
    = e^{i\spk_{\perp 2} \hat{\spx}_{\perp 2}} e^{i\hat{H}\tau}(e^{-i\hat{H}\tau})_{\hat{P}_3\rightarrow \hat{P}_3-k_3} e^{-i\spk_{\perp1} \hat{\spx}_{\perp1}},
\end{equation}
where $\tau:=t_2-t_1$. In order to proceed, we assume that
\begin{equation}\label{trans_coh0}
    e^{-i \spk_{\perp 1,2}\hat{\spx}_{\perp 1,2}}| i\ran\approx e^{-i \spk_{\perp 1,2}\spx_{\perp 1,2}}| i\ran,
\end{equation}
where $\spx_{1,2}$ is the average value of the corresponding operator with respect to the state $|i\ran$. The approximate equality takes place, if
\begin{equation}\label{trans_coh}
    k_\perp \s_\perp\ll1,
\end{equation}
where $\s_\perp$ is the characteristic transverse (with respect to the detector axis) size of the wave packet in the configuration space.

Notice that condition \eqref{trans_coh} also arises in considering the radiation of twisted photons by a bunch of charged particles \cite{BKb,bunches}. When condition \eqref{trans_coh} is satisfied, the probability of radiation of a twisted photon by the bunch of $N$ particles is the same as for one charged particle, multiplied by $N$ (incoherent radiation) or by $N^2$ (coherent radiation). To put it differently, the form of the wave packet does not affect the radiation spectrum of twisted photons in this case. It is clear that if \eqref{trans_coh} is violated, the approximation we use cannot be employed. In that case, the probability of radiation of twisted photons depends severely on the form of the wave packet and is not determined only by the average values of the particle momentum and coordinate. In particular, all the fine effects stemming from the form of the wave packet of a charged particle (see, e.g., \cite{Ivanov11,KotSerSchi,SeSuFr14,KarlJHEP,AngMackPiaz,BliokhVErev,LBThY}) cannot be reproduced by the semiclassical approach considered here. However, condition \eqref{trans_coh} cannot be strongly violated. It was shown in \cite{BKb,bunches} that, in a general position, the coherent radiation produced by a smooth wave packet of a charged particle is strongly suppressed for $k_\perp\s_\perp\gtrsim3$ due to destructive interference of the radiation amplitudes coming from different points of the wave packet.

Furthermore, we have
\begin{equation}
    |\lan[e^{-i \spk_{\perp 1,2}\hat{\spx}_{\perp 1,2}},\hat{\mathbf{P}}]\ran|/|\lan\hat{\mathbf{P}}^2\ran|^{1/2}\sim k_0/(\e\ga)\lesssim 1/\gamma,
\end{equation}
i.e., up to the terms of order $1/\gamma$, we can carry these exponents through the operators entering into $\hat{M}_2^\dag$ and $\hat{M}_1$ and make use of \eqref{trans_coh0}. Now take into account that
\begin{equation}\label{sqrt_app}
    \sqrt{(P_3-k_3)^2+P_\perp^2+m^2}\approx(P_0-k_0)\Big(1+\frac{k_0P_0-k_3P_3}{(P_0-k_0)^2}-\frac{k_\perp^2}{2(P_0-k_0)^2}\Big).
\end{equation}
The approximation \eqref{sqrt_app} is valid provided
\begin{equation}\label{recoil_est}
    \frac{P_0 k_0(1-\mathbf{n}\dot{\mathbf{x}})}{(P_0-k_0)^2}\sim\frac{k_0/P_0}{(1-k_0/P_0)^2}\frac{\vk^2}{\ga^2}\ll1.
\end{equation}
If $\vk/\gamma$ is small, estimate \eqref{recoil_est} holds up to $k_0\lesssim P_0$. Then, repeating the calculations presented in \cite{BaKaStrbook}, we can write
\begin{equation}\label{2exp1}
    e^{-ik_0 \tau} e^{i\hat{H}\tau} (e^{-i\hat{H}\tau})_{\hat{P}_3\rightarrow \hat{P}_3-k_3}\approx e^{-i\frac{\hat{P}_0}{\hat{P}_0-k_0} [k_0\tau-k_3(x_{32}-x_{31})-k_\perp^2\tau/(2\hat{P}_0)]},
\end{equation}
where we have used the fact that, in time-independent magnetic fields, $P_0=m\gamma$ is an integral of motion of the Lorentz equations. It is argued in \cite{BaKaStrbook} that approximation \eqref{2exp1} is valid in the leading order in $\vk/\gamma$ for general nonstationary external electromagnetic fields as well, provided
\begin{equation}
    2\pi/(T\e)\ll1,
\end{equation}
where $T$ is the characteristic time- or length-scale of changing of the external electromagnetic fields. In that case, $\hat{P}_0$ should be replaced by $\hat{P}_{0i}$ in \eqref{2exp1}, i.e., by the particle energy in the initial state $|i\ran$ where the external fields are absent. This prescription ensures, in particular, that the right-hand side of \eqref{2exp1} is invariant under translations in the spacetime.

Disentangling expression \eqref{2exp} and taking into account the conditions 1-4 and \eqref{ultrarel_appr}, \eqref{trans_coh}, we can replace the operators in matrix element \eqref{matr_elem0} by their average values. In particular,
\begin{equation}
    \hat{P}'_{0\,1,2}\rightarrow P'_{0\,1,2}=P_{0}(\mathbf{P}_{1,2}-\spk_{1,2})\approx P_{0\,1,2}-k_0.
\end{equation}
Then, in the scalar case,
\begin{equation}\label{matr_elem}
    C(\spk_2,\spk_1)=\frac{e^2}{(2\pi)^32Vk_0} c(\spk_2)c^*(\spk_1),
\end{equation}
where
\begin{equation}
    c(\spk):=\int_{-\infty}^\infty dt e^{-iq_i[k_0-k_\perp^2/(2P_{0i})]t+i q_i k_3x_{3} +i \spk_{\perp}\spx_{\perp}}
     q^{1/2} (\mathbf{f}\dot{\spx}),
\end{equation}
and
\begin{equation}
    \dot{\spx}(t)=\mathbf{P}(t)/P_0(t),\qquad q(t)=P_0(t)/P'_0(t),
\end{equation}
and $q_i=P_{0i}/P'_{0i}$. Recall that $P_{0i}$ is the energy of particle in the initial state and $P'_{0i}=P_{0i}-k_0$. The approximate expression for probability \eqref{prob_gen} derived from \eqref{matr_elem} is nonnegative. Furthermore, matrix element \eqref{matr_elem} with the photon momenta satisfying \eqref{phot_mom} transforms properly under translations in the spacetime, $x^\mu\rightarrow x^\mu+a^\mu$, viz.,
\begin{equation}\label{under_transl}
    C(\spk_2,\spk_1)\rightarrow C(\spk_2,\spk_1)e^{i(\spk_{2\perp}-\spk_{1\perp})\spa_\perp},
\end{equation}
where $\spa_\perp$ is the translation four-vector component perpendicular to the detector axis.

Employing formula \eqref{tw_w_in_pl_w}, we find the leading contribution to the probability of radiation of a twisted photon by a charged scalar particle with the quantum recoil taken into account
\begin{equation}\label{prob_by_scal}
    dP(s,m,k_\perp,k_3)=e^2 \Big|\int_{-\infty}^\infty dt e^{-iq_i[k_0-k_\perp^2/(2P_{0i})]t+i q_i k_3x_{3}}
    q^{1/2} \big(\tfrac12\big[\dot{x}_{+} a_- +\dot{x}_{-} a_+ \big] +\dot{x}_{3} a_3 \big)\Big|^2 \Big(\frac{k_\perp}{2k_{0}}\Big)^{3}\frac{dk_3dk_\perp}{2\pi^2},
\end{equation}
where
\begin{equation}
    a_\pm\equiv a_\pm(s,m,k_3,k_\perp; \spx ),\qquad a_3\equiv a_3 (s,m,k_\perp; \spx).
\end{equation}
Recall that we use the system of units such that $e^2=4\pi\al$. In the case $q\approx1$, i.e., when the energy of radiated photon is negligibly smaller than the energy of charged particle, formula \eqref{prob_by_scal} passes into formula [(36), \cite{BKL2}]. As a rule, $q\approx const$ in the ultrarelativistic limit and so $q^{1/2}$ can be removed from the integrand of \eqref{prob_by_scal}. Since $q>1$, we see that the quantum recoil tends to increase the radiation probability in comparison with the answer which does not include it.


The estimate \eqref{trans_coh} is most stringent condition on the range of applicability of formula \eqref{prob_by_scal}. It is absent in the semiclassical description of radiation of plane wave photons \cite{BaKaStrbook}. The narrower transverse size of a particle wave packet, the better this condition is satisfied. Currently, there are techniques allowing to produce the electron wave packets with the waist of order $1$ {\AA} \cite{Vereeck11}. Theoretically, for a given instant of time, the electron wave packet can be focused into a region much smaller than the Compton wavelength. However, the problem is how to create such wave packets and keep them in this state for a sufficiently long time so that the electron will have time to radiate a twisted photon. For a particle, which is at rest on average, the width of the wave packet cannot be smaller than the Compton wavelength. Otherwise, the momentum uncertainty is larger than $2m$ and the height of a potential barrier confining such a particle must be larger than $2m$. Such fields create electron-positron pairs, and the problem becomes essentially multiparticle. Nevertheless, if the electron is moving on average, then its wave packet can be squeezed stronger in the transverse directions with the aid of the fields not exceeding the critical (Schwinger) field in the laboratory frame. In the comoving reference frame, the potential well depth can become larger than the critical one but this does not result in pair creation. In passing to the comoving frame, the magnetic field arises and the respective invariants of the electromagnetic field remain the same as in the laboratory frame.

This situation is naturally realized for axial channeling of particles in crystals (see for details, e.g., \cite{KimCue,AkhShul,BaKaStrbook,Bord.1,Baryshev}). Suppose that the particle momentum component $p_3\gg m$ and $|p_{1,2}|\ll p_3$. Then the solution of the stationary Dirac or Klein-Gordon equations is reduced approximately to the solution of the Schr\"{o}dinger equation with a certain effective potential $U(\spx_\perp)$:
\begin{equation}
    \Big(\frac{\spp_\perp^2}{2 m_{eff}}+U\Big)\psi=\de E\psi,
\end{equation}
where $\de E=E-m_{eff}$, $m_{eff}^2:=m^2+p_3^2$, and it is assumed that $|U|\ll m_{eff}$ and $|\de E|\ll m_{eff}$. The effective mass is well approximated by
\begin{equation}
    m_{eff}\approx \ga m.
\end{equation}
The potential barrier of the height $U_0$ can hold a particle with typical momentum uncertainty
\begin{equation}
    |\De p_\perp|\lesssim \sqrt{2 m_{eff} U_0}.
\end{equation}
Then from the uncertainty relation, we obtain the minimum transverse size
\begin{equation}
    \s_\perp^m\sim (2 m_{eff} U_0)^{-1/2}\approx1/(\theta_c\e),
\end{equation}
where the critical Lindhard angle has been introduced
\begin{equation}
    \theta_c:=\sqrt{\frac{2 U_0}{\e}},\qquad U_0\sim Z\al^2 m.
\end{equation}
Since $\be_\perp\approx\theta_c$, then
\begin{equation}\label{sigma_perp_m}
    \s_\perp^m\sim 1/(m K).
\end{equation}
As we see, the transverse size of the particle wave packet can be much smaller than the Compton wavelength for axal channeling in the wiggler regime \cite{AugSchGrei,OlsKun}. The maximum transverse squeezing of a particle wave packet is achieved at
\begin{equation}
    \s_\perp^m\sim1/(m\ga).
\end{equation}
However, this is possible only in the external fields of the same order as the critical one in the laboratory frame. Such maximally localized wave packets were investigated in \cite{BilBer17,BilBer19}.

Consider the fulfillment of condition \eqref{trans_coh} for axial channeling with the wave packet transverse size \eqref{sigma_perp_m}. The respective undulator frequency is estimated as
\begin{equation}
    \omega\sim\frac{\pi\theta_c}{d},
\end{equation}
where $d$ is the channel width,
\begin{equation}
    d\sim 1\,\text{{\AA}}\sim 10^3 m^{-1}.
\end{equation}
Then, in the dipole regime, at the $n$-th harmonic
\begin{equation}
    k_0\sim 2\omega\ga^2 n.
\end{equation}
Hence,
\begin{equation}
    k_\perp\s_\perp^m\sim 2\pi\frac{n_\perp\ga}{md}n\ll1.
\end{equation}
Bearing in mind that $n_\perp\ga\lesssim1$, we see that \eqref{trans_coh} is satisfied. This condition is fulfilled even for $\s_\perp=d$ in the dipole regime for not too large harmonic numbers. In the wiggler case for $n_\perp\ga\approx K$, we have
\begin{equation}
    k_0\sim \omega\ga^2 n/K^2.
\end{equation}
Then
\begin{equation}
    k_\perp\s_\perp^m\sim\frac{\pi n}{mdK}\ll1.
\end{equation}
In fact, condition \eqref{trans_coh} is already satisfied for $\s_\perp n\lesssim d/10$. The energy shift due to quantum recoil was neglected in these formulas. The inclusion of quantum recoil only improves the estimate.

The dynamical picture looks as follows. The wide particle wave packet falls onto the crystal surface and, in the channeling regime, is split into wave packets with transverse sizes from \eqref{sigma_perp_m} up to $d$. Because of interaction with the crystalline potential and other electrons, the different parts of the wave packet moving in different channels do not almost interfere. In order to describe the radiation of twisted photons produced by each part of such dispersed wave packet, formula \eqref{prob_by_scal} can be used or, rather, its analog for Dirac particles that will be obtained below. As the particle escaped the crystal, the parts of the wave packet spread in the transverse directions with the characteristic velocity $\be_\perp\approx\theta_c$.

\paragraph{Dirac particle.}

In case of radiation of twisted photons by the Dirac particles, the considerations are analogous but more cumbersome. Using the notation of \cite{BaKaStrbook}, we have approximately
\begin{equation}
\begin{split}
    M(t_1,\spk_1) &= \Big(\frac{m}{P_{01}}\Big)^{1/2} \bar{u}_{s'}(\mathbf{P}_1)\hat{\mathbf{f}}^*_1e^{-i\spk_1\spx_1} u_{s}(\mathbf{P}_1) \Big(\frac{m}{P_{01}}\Big)^{1/2}= e^{-i\spk_1\spx_1} \Big(\frac{m}{P'_{01}}\Big)^{1/2} \bar{u}_{s'}(\mathbf{P}'_1) \hat{\mathbf{f}}^*_1 u_{s}(\mathbf{P}_1) \Big(\frac{m}{P_{01}}\Big)^{1/2},\\
    M^\dag(t_2,\spk_2) &= \Big(\frac{m}{P_{02}}\Big)^{1/2} \bar{u}_{s}(\mathbf{P}_2)\hat{\mathbf{f}}_2 e^{i\spk_2\spx_2} u_{s'}(\mathbf{P}_2) \Big(\frac{m}{P_{02}}\Big)^{1/2}= \Big(\frac{m}{P_{02}}\Big)^{1/2} \bar{u}_{s}(\mathbf{P}_2) \hat{\mathbf{f}}_2 u_{s'}(\mathbf{P}'_2) \Big(\frac{m}{P'_{02}}\Big)^{1/2} e^{i\spk_2\spx_2},
\end{split}
\end{equation}
where $s$ and $s'$ characterize the initial and final electron spin states. For brevity, we do not write hats over the operators $\mathbf{P}$ and $\spx$ anymore. Further, we employ formulas \eqref{2exp}, \eqref{2exp1}, substitute all the operators by their averages, sum over spin polarizations of the escaping electron and average over spin polarizations of the incoming electron. As a result, we come to
\begin{equation}
    \frac12\sum_{\text{spins}}M^\dag(t_2,\spk_2) M(t_1,\spk_1)\rightarrow\frac12\Sp\big[(A^*_2-i(\mathbf{B}^*_2 \boldsymbol{\s})) (A_1+i(\mathbf{B}_1 \boldsymbol{\s})) \big]=A^*_2A_1+(\mathbf{B}^*_2\mathbf{B}_1),
\end{equation}
where
\begin{equation}\label{A_B}
\begin{split}
    A_{1,2}&=-\frac{(\mathbf{f}^*_{1,2}\mathbf{P}_{1,2})}{2\sqrt{P'_{0\,1,2} P_{0\,1,2}}} \bigg(\sqrt{\frac{P'_{0\,1,2}+m}{P_{0\,1,2}+m}} +\sqrt{\frac{P_{0\,1,2}+m}{P'_{0\,1,2}+m}}\bigg),\\
    \mathbf{B}_{1,2}&=-\frac{1}{2\sqrt{P'_{0\,1,2} P_{0\,1,2}}} \bigg(\sqrt{\frac{P'_{0\,1,2}+m}{P_{0\,1,2}+m}} [\mathbf{f}^*_{1,2},\mathbf{P}_{1,2}] -\sqrt{\frac{P_{0\,1,2}+m}{P'_{0\,1,2}+m}} [\mathbf{f}^*_{1,2},\mathbf{P}'_{1,2}] \bigg).
\end{split}
\end{equation}
In the ultrarelativistic limit, we can neglect the mass in expressions \eqref{A_B} since it gives the contributions of order $1/\gamma$ as compared to the main contribution. Then
\begin{equation}\label{AA_BB}
\begin{split}
    A^*_2A_1+(\mathbf{B}^*_2\mathbf{B}_1)&= \frac{1}{4 P'_{01} P'_{02}}\big[ (P_{01}+P_{01}')(P_{02}+P_{02}')(\mathbf{f}_2\dot{\mathbf{x}}_2)(\mathbf{f}^*_1\dot{\mathbf{x}}_1) -\\
    &-k_0^2 (\mathbf{f}_2,\dot{\mathbf{x}}_1-\mathbf{n}_1)(\mathbf{f}^*_1,\dot{\mathbf{x}}_2-\mathbf{n}_2) +k_0^2(\mathbf{f}_2\mathbf{f}^*_1)(\dot{\mathbf{x}}_2-\mathbf{n}_2,\dot{\mathbf{x}}_1-\mathbf{n}_1) \big].
\end{split}
\end{equation}
This expression turns into formula [(2.41), \cite{BaKaStrbook}] for $\spk_{1,2}=\spk$.

The integration over azimuth angles of the vectors $\spk_{1,2}$ is considered in Appendix \ref{Azim_Int}. With the aid of the notation introduced there, we can write
\begin{equation}\label{prob_by_Dir}
\begin{split}
    dP(s,m,k_\perp,k_3)&=e^2\int_{-\infty}^\infty \frac{dt_1dt_2}{4P'_{01} P'_{02}} e^{-i(k_0-k_\perp^2/(2P_{0i}))q_i(t_2-t_1)+ik_3q_i(x_{23}-x_{13})}\times\\
    &\times\Big\{(P_{01}+P'_{01})(P_{02}+P'_{02})(\tfrac12[\dot{x}_{1-} a^*_- +\dot{x}_{1+} a^*_+]+\dot{x}_{13}a^*_3) (\tfrac12[\dot{x}_{2+} a_- +\dot{x}_{2-} a_+]+\dot{x}_{23}a_3)+\\
    &+\frac{k_0^2}{4}\big[(\dot{x}_{1+}a^*_+-in_\perp a^*_+(m-1)) (\dot{x}_{2-}a_+ +in_\perp a_+(m-1))+ \\
    &+(\dot{x}_{1-} a^*_-+in_\perp a^*_-(m+1)) (\dot{x}_{2+}a_--in_\perp a_-(m+1)) \big]\Big\}\Big(\frac{k_\perp}{2k_{0}}\Big)^{3}\frac{dk_3dk_\perp}{2\pi^2}
\end{split}
\end{equation}
in the leading order in $\vk/\gamma$. Here
\begin{equation}\label{a_pm3}
\begin{gathered}
    a_\pm\equiv a_\pm(s,m,k_3,k_\perp; \spx_2 ),\qquad a_3\equiv a_3 (s,m,k_\perp; \spx_2),\\
    a^*_\pm\equiv a^*_\pm(s,m,k_3,k_\perp; \spx_1),\qquad a^*_3\equiv a^*_3(s,m,k_\perp; \spx_1).
\end{gathered}
\end{equation}
The approximate expression \eqref{prob_by_Dir} for probability density \eqref{prob_gen} is nonnegative. In the case $k_0\ll\e_{1,2}$, formula \eqref{prob_by_Dir} passes into expression [(36), \cite{BKL2}] without the quantum recoil due to photon radiation. Just as for the scalar particle case, we see that the quantum recoil tends to increase the radiation probability as compared to the formula without recoil. As long as
\begin{equation}
    \frac{P_0+P_0'}{2P_0'}=\frac{1+q}{2},\qquad \frac{k_0}{P_0'}=q-1=:\de q,
\end{equation}
the second and third terms in \eqref{prob_by_Dir} are proportional to $(\de q)^2$, while the first term is proportional to $(1+\de q/2)^2$. Thus, in the limit of small recoil, $q\approx 1$, the second and third terms can be neglected and the contribution of the first term is the same as in the case of a scalar particle \eqref{prob_by_scal} since $q^{1/2}\approx 1+\de q/2$. The spin effects become irrelevant in this limit within the bounds of the approximations made in deriving \eqref{prob_by_Dir}.

Contrary to the classical formula for radiation probability, expression \eqref{prob_by_Dir} does not factorize into $c^*c$, where $c$ is determined by the particle trajectory. This is a consequence of the fact that the averaging over spin polarizations was performed and the quantum recoil was taken into account. If one neglects the recoil, the created radiation will be described by a coherent state in the Fock space \cite{GottfYan,KlauSud,Glaub1,Glaub2,BKL2} and nontrivial quantum correlations will be absent. In the case of a scalar charged particle, formula \eqref{prob_by_scal} does factorize into the amplitude and its complex conjugate, and the nontrivial quantum correlations are absent within the bounds of the approximations made.

Notice that formulas \eqref{prob_by_scal}, \eqref{prob_by_Dir} describe the radiation produced by one charged particle. As a rule, in real experiments, the bunch of charged particles radiates. Under usual conditions, radiation amplitudes of hard photons produced by different particles in the bunch add up incoherently, i.e., expressions \eqref{prob_by_scal} or \eqref{prob_by_Dir} should be summed over different particles in the bunch. In the papers \cite{BKb,bunches}, the simple formulas were obtained allowing to find the radiation of twisted photons by a bunch of charged particles using the one-particle radiation probability distribution. These formulas can be applied to \eqref{prob_by_scal} and \eqref{prob_by_Dir} as the initial plane wave matrix elements \eqref{under_transl} transform correctly under translations. Below, we shall employ the formulas from \cite{BKb} for axially symmetric bunches. Such bunches are created, for example, in the electron-positron collider VEPP-2000, Novosibirsk \cite{RPPPDG2018}.

\section{Forward radiation by charged particles in undulators}\label{Undul}

Let us apply the general formulas derived in the preceding section to the forward radiation of charged particles in undulators with the quantum recoil taken into account. We shall investigate the undulator radiation in the dipole regime for an arbitrary periodic trajectory of a charged particle. As for the wiggler case, we shall obtain the exact expression for the probability of radiation of twisted photons by charged particles moving along an ideal helix (the helical wiggler).

\paragraph{Scalar particle.}

The general formula for the radiation probability of a twisted photon by a scalar charged particle has the form \eqref{prob_by_scal}. The radiation of twisted photons by undulators without quantum recoil was described in (\cite{BKL2}, Sec. 5).
Comparing [(82), \cite{BKL2}] with \eqref{prob_by_scal}, we see that the account for quantum recoil leads only to a change of the energy spectrum of radiated photons and to the appearance of the common factor $q$ when the forward radiation produced by a scalar particle in the undulator in the dipole regime and in the ideal helical wiggler is considered. Notice that, according to the Lorentz equations,
\begin{equation}
    q=q_i=const
\end{equation}
for the motion of charged particles in undulators. As a result, in the dipole approximation,
\begin{equation}
\begin{split}
    dP(s,m,k_\perp,k_3)&=e^2 n_\perp^3\sum_{n=1}^\infty\de_N^2\Big[qk_0\Big(1-n_3\ups_3-\frac{n_\perp k_\perp}{2P_0}\Big) -n\omega \Big]\times\\
    &\times q\Big\{\de_{m,1}\Big(k_\perp\ups_3 +\frac{\omega nn_\perp}{n_3-s}\Big)^2|r_+(n)|^2 +\de_{m,-1}\Big(k_\perp\ups_3 +\frac{\omega nn_\perp}{n_3+s}\Big)^2|r_-(n)|^2\Big\}\frac{dk_3 dk_\perp}{16},
\end{split}
\end{equation}
where the notation introduced in \cite{BKL2} was used. As in the case of radiation without recoil, the main part of forward radiation consists of twisted photons with $m=\pm1$.

The energy spectrum is found from the equation
\begin{equation}\label{spectrum}
    P_0 k^n_0\Big(1-n_3\ups_3-\frac{n^2_\perp k_0^n}{2P_0}\Big) =n\omega(P_0-k_0^n).
\end{equation}
In virtue of condition \eqref{trans_coh} with minimum $\s_\perp$ taken from \eqref{sigma_perp_m}, the last term in the round brackets on the left-hand side is small and should be taken into account perturbatively. Indeed,
\begin{equation}
    1-n_3\ups_3-\frac{n^2_\perp k_0^n}{2P_0}\approx\frac{1+K^2+n_\perp^2\ga^2-n_\perp^2\ga^2 k_0^n/P_0}{2\ga^2}=\frac{1+K^2+n_\perp^2\ga^2/q}{2\ga^2}.
\end{equation}
For $k_0\sim\e$, it follows from \eqref{trans_coh}, \eqref{sigma_perp_m} that $n_\perp\ga\ll K$, and the contribution of this term can be neglected as compared to the contributions of the first terms. For $k_0\ll\e$, obviously, the contribution of this term is also negligibly small in comparison with the contributions of the first terms in this expression. Then the physical solution to \eqref{spectrum} takes the form
\begin{equation}\label{spectrum_sol}
\begin{split}
    k_0^n&=\frac{P_0(1-n_3\ups_3)}{n_\perp^2}\Big\{1+\frac{\bar{k}^n_0}{P_0}-\Big[\Big(1+\frac{\bar{k}^n_0}{P_0}\Big)^2 -\frac{2\bar{k}_0^n n_\perp^2}{P_0(1-n_3\ups_3)}\Big]^{1/2} \Big\}\approx\\
    &\approx \frac{P_0(1+K^2+n_\perp^2\ga^2)}{2n_\perp^2\ga^2}\Big\{1+\frac{\bar{k}^n_0}{P_0}-\Big[\Big(1+\frac{\bar{k}^n_0}{P_0}\Big)^2 -\frac{4\bar{k}_0^n n_\perp^2\ga^2}{P_0(1+K^2+n_\perp^2\ga^2)}\Big]^{1/2} \Big\},
\end{split}
\end{equation}
where
\begin{equation}
    \bar{k}_0^n:=\frac{\omega n}{1-n_3\ups_3}
\end{equation}
is the energy of radiated twisted photon without quantum recoil. If condition \eqref{trans_coh} with minimum $\s_\perp$ taken from \eqref{sigma_perp_m} is fulfilled, the last term under the square root in \eqref{spectrum_sol} is small. Developing \eqref{spectrum_sol} as a series, we obtain in the leading order
\begin{equation}\label{spectrum_sol_app}
    k_0^n=\frac{n\omega}{1-n_3\ups_3+n\omega/P_0}=\frac{\bar{k}_0^n}{1+\bar{k}_0^n/P_0}\;\Leftrightarrow\; \frac{1}{k_0^n}=\frac{1}{\bar{k}_0^n}+\frac{1}{P_0}.
\end{equation}
In fact, this is the BK prescription for the shift of the energy spectrum due to quantum recoil. Formulas \eqref{spectrum_sol}, \eqref{spectrum_sol_app} imply that $k_0^n<P_0\approx\e$.

Using formulas from \cite{BKL2}, the radiation of twisted photons by a scalar charged particle moving along an ideal helix can readily be derived. Denoting by $\chi=\pm1$ the helicity of the particle trajectory, we find from \eqref{prob_by_scal} that
\begin{equation}\label{wiggler_scal}
\begin{split}
    dP(s,m,k_\perp,k_3)&=e^2 n_\perp^3 \de_N^2\Big[qk_0\Big(1-n_3\ups_3-\frac{n_\perp k_\perp}{2P_0}\Big) -\chi m\omega \Big]\times\\
    &\times q\Big[\Big(\ups_3-\chi\frac{n_3\omega m}{n_\perp k_\perp}\Big)J_m\Big(\frac{k_\perp K}{\omega\ga}\Big) -\chi\frac{sK}{n_\perp\ga}J'_m\Big(\frac{k_\perp K}{\omega\ga}\Big) \Big]^2 \frac{dk_3 dk_\perp}{4}.
\end{split}
\end{equation}
Thus the selection rule $m=\chi n$ for the forward radiation produced by an ideal helical wiggler survives even when the quantum recoil is taken into account, within the bounds of the approximations made in deriving formula \eqref{prob_by_scal}. The energy spectrum of radiated twisted photons is given by \eqref{spectrum_sol}, \eqref{spectrum_sol_app}.

\paragraph{Dirac particle.}

In the dipole approximation, it is necessary to expand the integrand of \eqref{prob_by_Dir} into series in $K$ and take into account only the leading contribution (see the estimates in Sec. 5.A of \cite{BKL2}). Then, up to a common factor, the first term in the curly brackets in \eqref{prob_by_Dir} turns into the same expression as that appearing when the quantum recoil is neglected,
\begin{equation}
\begin{split}
    \frac14(P_0+P'_0)^2\Big[&-\de_{m,1}\Big(\frac{in_\perp}{n_3-s}\dot{r}_{2+} +k_\perp\ups_3 r_{2+}\Big) \Big(\frac{in_\perp}{n_3-s}\dot{r}_{1-}+k_\perp\ups_3 r_{1-}\Big)-\\
    &-\de_{m,-1}\Big(\frac{in_\perp}{s+n_3}\dot{r}_{2-} -k_\perp\ups_3 r_{2-}\Big) \Big(\frac{in_\perp}{s+n_3}\dot{r}_{1+}+k_\perp\ups_3 r_{1+}\Big) \Big].
\end{split}
\end{equation}
The second term in the curly brackets in \eqref{prob_by_Dir} standing at $k_0^2/4$ is written as
\begin{equation}
\begin{split}
    \frac{n_\perp^2}{(s+n_3)^2} &\Big[\dot{r}_{1+}\de_{m,-1}-in_\perp\big(\de_{m,0} +\tfrac12 \de_{m,1}k_\perp r_{1-} -\tfrac12 \de_{m,-1}k_\perp r_{1+} \big) \Big]\times\\
    \times&\Big[\dot{r}_{2-}\de_{m,-1}+in_\perp\big(\de_{m,0} +\tfrac12 \de_{m,1}k_\perp r_{2+} -\tfrac12 \de_{m,-1}k_\perp r_{2-} \big) \Big]+\\
    +\frac{n_\perp^2}{(s-n_3)^2} &\Big[\dot{r}_{1-}\de_{m,1}+in_\perp\big(\de_{m,0} +\tfrac12 \de_{m,1}k_\perp r_{1-} -\tfrac12 \de_{m,-1}k_\perp r_{1+} \big) \Big]\times\\
    \times&\Big[\dot{r}_{2+}\de_{m,1}-in_\perp\big(\de_{m,0} +\tfrac12 \de_{m,1}k_\perp r_{2+} -\tfrac12 \de_{m,-1}k_\perp r_{2-} \big) \Big].
\end{split}
\end{equation}
Expanding $r_{1,2}$ into a Fourier series and integrating over time, we have
\begin{multline}
    dP(s,m,k_\perp,k_3)=e^2 n_\perp^3\sum_{n=1}^\infty\de_N^2\Big[qk_0\Big(1-n_3\ups_3-\frac{n_\perp k_\perp}{2P_0}\Big) -n\omega \Big]\times\\
    \times\Big\{\de_{m,1}\Big[(P_0+P_0')^2\Big(k_\perp\ups_3+\frac{\omega nn_\perp}{n_3-s}\Big)^2 +\frac{k_\perp^2}{(n_3-s)^2}\Big(\omega n -\frac{n_\perp k_\perp}{2}\Big)^2 +\frac{n_\perp^2 k_\perp^4}{4 (n_3+s)^2} \Big] |r_+(n)|^2+\\
    +\de_{m,-1}\Big[(P_0+P_0')^2\Big(k_\perp\ups_3+\frac{\omega nn_\perp}{n_3+s}\Big)^2 +\frac{k_\perp^2}{(n_3+s)^2}\Big(\omega n -\frac{n_\perp k_\perp}{2}\Big)^2 +\frac{n_\perp^2 k_\perp^4}{4 (n_3-s)^2} \Big] |r_-(n)|^2 \Big\}\frac{dk_3 dk_\perp}{64 P_0'^2}.
\end{multline}
As in the case of a scalar particle, only the twisted photons with $m=\pm1$ are radiated. The photon energy spectrum has the form \eqref{spectrum_sol}, \eqref{spectrum_sol_app}.

\begin{figure}[t]
   \centering
\includegraphics*[align=c,width=0.47\linewidth]{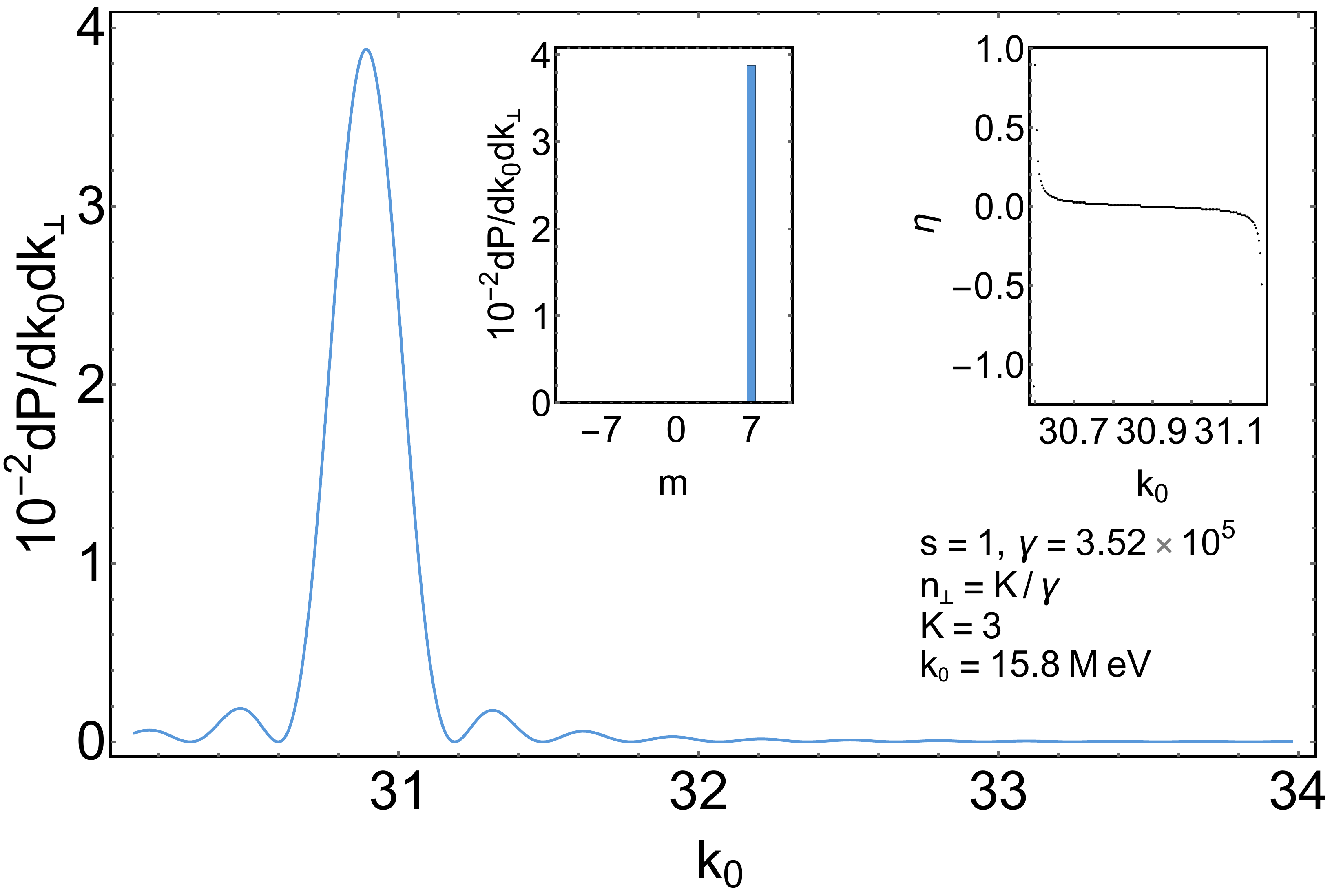}\;
\includegraphics*[align=c,width=0.48\linewidth]{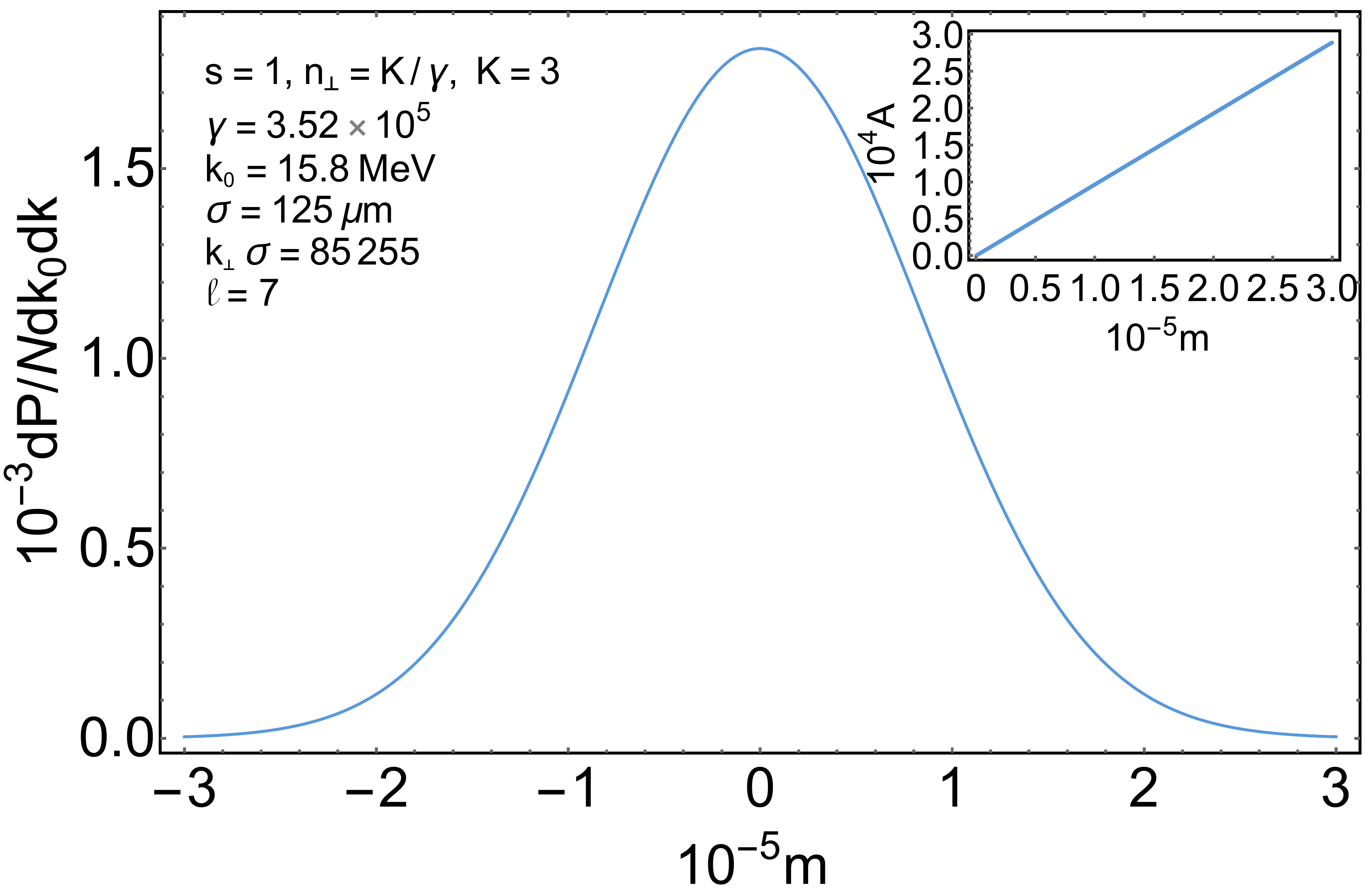}
    \caption{{\footnotesize The radiation of twisted photons by $180$ GeV electrons in the helical wiggler. The seventh harmonic is presented. The wiggler period $0.72$ cm, the number of periods $N=15$, and the magnetic field strength in the wiggler $63.3$ kG. The applicability conditions \eqref{m_ineq} are satisfied for $\s_\perp\lesssim10^3m^{-1}$. The photon energy is measured in the electron rest energies. Left panel: The probability to record a twisted photon produced by one electron in the wiggler against the photon energy. Left inset: The probability distribution over $m$. Right inset: The relative change of radiation probability due to quantum recoil: $\eta:=(dP_{cl}-dP)/dP_{cl}$, where $dP_{cl}$ is the radiation probability without quantum recoil. Right panel: The distribution over $m$ of probability per particle to record a twisted photon produced by an incoherent axially symmetric bunch of particles in the wiggler. The bunch width is $\s=125\mu$m. The angular momentum per photon is the same as for radiation of one electron. Inset: The asymmetry of distribution over $m$.}}
    \label{fig_WIGG}
\end{figure}

\begin{figure}[t]
   \centering
\includegraphics*[align=c,width=0.48\linewidth]{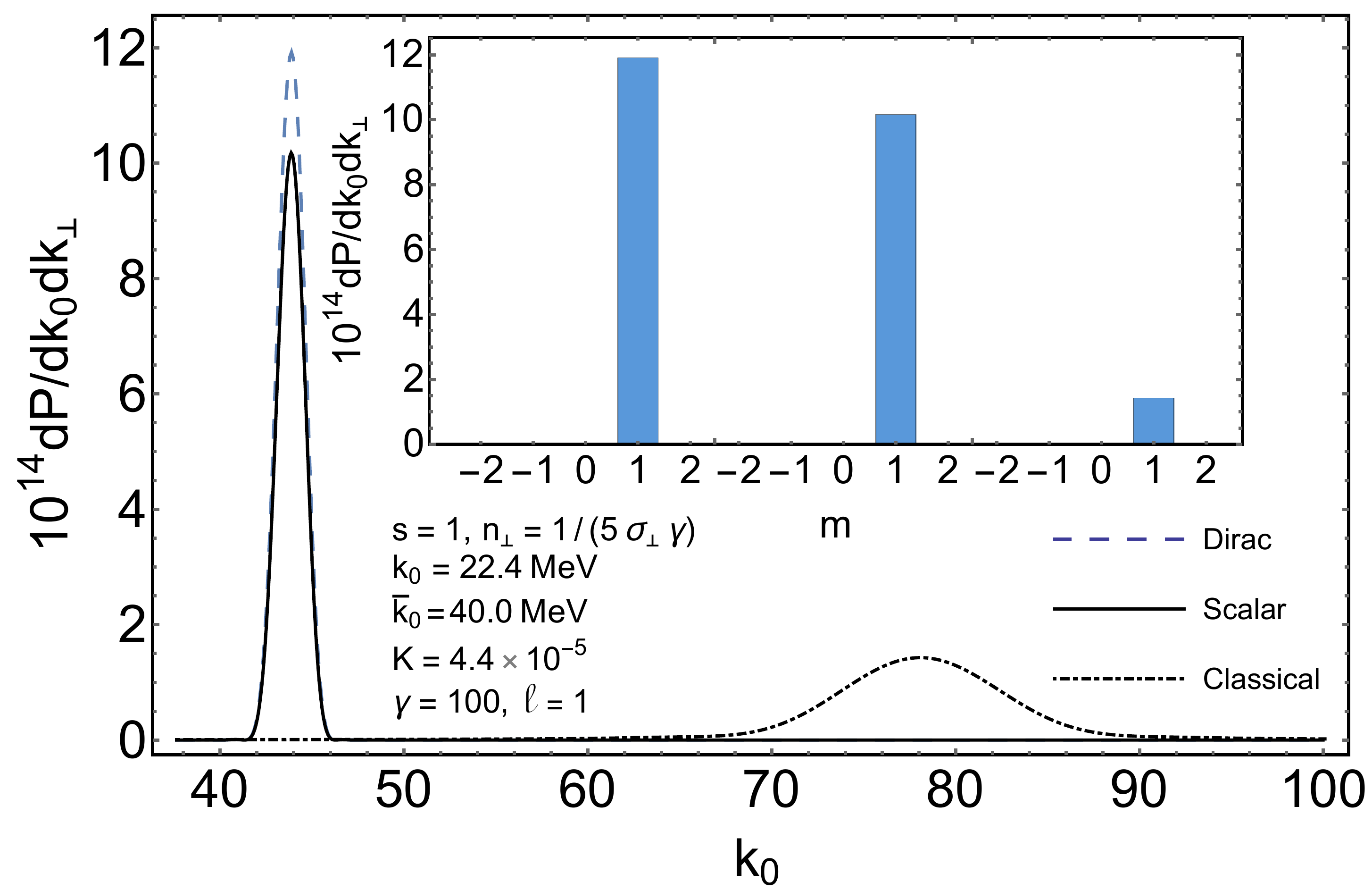}\;
\includegraphics*[align=c,width=0.48\linewidth]{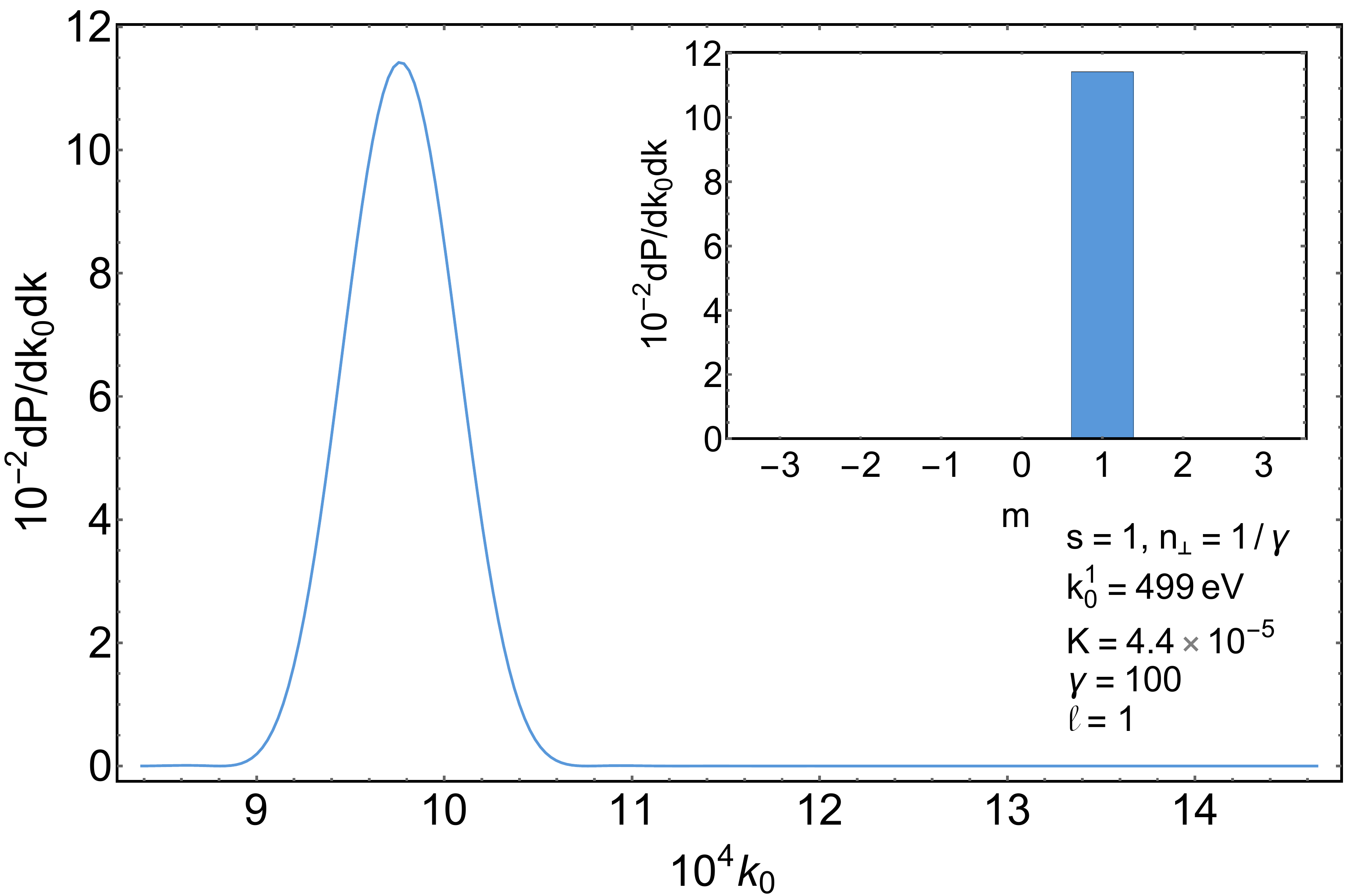}
    \caption{{\footnotesize The radiation of twisted photons by $51.1$ MeV electrons evolving in the circularly polarized  electromagnetic wave produced by the free electron laser with the photon energy $1$ keV, intensity $3.38\times10^15$ W/cm${}^2$, and amplitude envelope \eqref{envelope} with $N=20$. These data correspond to $\Omega\approx1.96\times 10^{-3}$ and $a_0\approx8.54\times 10^{-8}$. The applicability conditions \eqref{m_ineq} are satisfied for $\s_\perp\lesssim10^3m^{-1}$. The photon energy is measured in the electron rest energies. Left panel: The head-on collision. The first harmonic \eqref{harmonics_new} for radiation without recoil (classical), scalar, and Dirac particles is depicted. For the parameters chosen, the quantum recoil halves the energy of radiated photons in comparison with formula without recoil. The probability of radiation of twisted photons by Dirac particles is bigger than by the one by scalar particles which, in turn, is bigger than the probability of radiation of twisted photons without quantum recoil (see the discussion after Eq. \eqref{prob_by_Dir_wigg}). Right panel: The laser wave is overtaking the electron. The first harmonic \eqref{harmonics_new} is shown. The quantum recoil is negligible in this case. Insets: The distributions over $m$ at the main maxima of harmonics.}}
    \label{fig_FEL}
\end{figure}

Performing calculations along the same lines as in the case of a scalar particle, we find for an ideal helical wiggler
\begin{multline}\label{hel_wigg_D}
    dP(s,m,k_\perp,k_3)=e^2 n_\perp^3 \de_N^2\Big[qk_0\Big(1-n_3\ups_3-\frac{n_\perp k_\perp}{2P_0}\Big) -\chi m\omega \Big]\times\\
    \times \Big\{(P_0+P_0')^2\Big[\Big(\ups_3 -\chi\frac{n_3\omega m}{n_\perp k_\perp}\Big)J_m -\chi\frac{sK}{n_\perp\ga}J'_m \Big]^2 +\frac{k_\perp^2}{4(n_3+s)^2}\Big(\frac{K}{\ga}J_{m+1} -\chi n_\perp J_m\Big)^2+\\ +\frac{k_\perp^2}{4(n_3-s)^2}\Big(\frac{K}{\ga}J_{m-1} -\chi n_\perp J_m\Big)^2 \Big\} \frac{dk_3 dk_\perp}{16P_0'^2}.
\end{multline}
The argument of the Bessel functions is the same as in formula \eqref{wiggler_scal}. Employing the recurrence relations for the Bessel functions and keeping in mind that $n_3\approx1$, the last two terms in the curly brackets can be brought to
\begin{equation}
    k_0^2\Big[\Big(\frac{\omega m}{n_\perp k_\perp}-1\Big)J_m +\frac{sK}{n_\perp\ga}J'_m\Big]^2,
\end{equation}
to the accuracy we work. Setting $\ups_3=n_3=1$ in the first term in the curly brackets in \eqref{hel_wigg_D}, we obtain
\begin{equation}\label{prob_by_Dir_wigg}
\begin{split}
    dP(s,m,k_\perp,k_3)&\approx e^2 n_\perp^3 \de_N^2\Big[qk_0\Big(1-n_3\ups_3-\frac{n_\perp k_\perp}{2P_0}\Big) -\chi m\omega \Big]\times\\
    &\times (1+q^2) \Big[\Big(1 -\chi\frac{\omega m}{n_\perp k_\perp}\Big)J_m -\chi\frac{sK}{n_\perp\ga}J'_m \Big]^2 \frac{dk_3 dk_\perp}{8}.
\end{split}
\end{equation}
The forward radiation of twisted photons in helical wigglers obeys the selection rule $m=\chi n$ within the bounds of the approximations made (see Fig. \ref{fig_WIGG}). Comparing \eqref{wiggler_scal} with \eqref{prob_by_Dir_wigg}, we see that the probability of radiation of twisted photons by Dirac particles is always bigger than the same quantity for scalar particles since $(1+q^2)/2-q>0$ for $q>1$ (see Fig. \ref{fig_FEL}). For $q\approx1$, the respective radiation probabilities are almost equal.

\paragraph{Number of radiated twisted photons.}

Let us find the number of radiated twisted photons with a given projection of the total angular momentum $m$ per energy interval $k_0$ and the total number of radiated twisted photons. Further, we set $\chi=1$ since $\chi=-1$ can be obtained by the substitution $m\rightarrow-m$, $s\rightarrow-s$. To shorten formulas, we suppose that the last term in the round brackets in \eqref{spectrum} is small and the energy spectrum is given by \eqref{spectrum_sol_app}. Then
\begin{equation}\label{meas_trans}
    \de_N^2[qk_0 (1-n_3\ups_3) -m\omega]dk_3 dk_\perp=\de_N^2(x)\frac{dk_0 dx}{\ups_3 q n_\perp\omega}\approx \de_N^2(x)\frac{dk_0 dx}{q n_\perp\omega}.
\end{equation}
As long as the function $\de_N^2(x)$ is localized near the point $x=0$ for $N\gtrsim 5$, the integral over $x$ can easily be found
\begin{equation}\label{int_deN}
    \int_{-\infty}^\infty dx \de_N^2(x)=N.
\end{equation}
Substituting \eqref{meas_trans}, \eqref{int_deN} into \eqref{wiggler_scal}, \eqref{prob_by_Dir_wigg}, we obtain $dP(s,m,k_0)/dk_0$. The analogous formulas can be obtained in the dipole case as well. However, in that case, the overlapping of harmonics has to be taken into account. Namely, the integration over $x$ results in
\begin{equation}
    \sum_{n=1}^\infty\de_N^2[qk_0(1-n_3\ups_3) -n\omega ]dk_3 dk_\perp\rightarrow N\sum_{n=1}^\infty \theta_{k_0}\Big[\frac{n\omega}{1+n\omega/P_0},\frac{n\omega}{1-\ups_3+n\omega/P_0}\Big]\frac{dk_0}{q n_\perp\omega},
\end{equation}
where
\begin{equation}
    \theta_{x}[a,b]:=\left\{
                        \begin{array}{ll}
                          1, & \hbox{$x\in[a,b]$;} \\
                          0, & \hbox{$x\not\in[a,b]$,}
                        \end{array}
                      \right.
\end{equation}
and $n_\perp$ is to be expressed through $k^n_0$ by using \eqref{spectrum_sol_app}.

The twisted photons with large projections $m$ of the total angular momentum can be produced by undulators only in the wiggler regime. When $m\gtrsim 5$, the Bessel functions entering into \eqref{wiggler_scal}, \eqref{prob_by_Dir_wigg} are expressed through the Airy functions \cite{Bord.1} with (see the notation in [(122), \cite{BKL2}])
\begin{equation}\label{x_Airy}
    x\approx1-\frac{4n_\perp^2\ga^2 K^2}{(1+K^2+n_\perp^2\ga^2)^2q^2}.
\end{equation}
For the probability of radiation of twisted photons with the total angular momentum projection $m$, $m\gtrsim5$, not to be exponentially suppressed, quantity \eqref{x_Airy} must be small. This is achieved when $K\gtrsim3$ and
\begin{equation}\label{n_k_def}
    n_k:=n_\perp \ga/K\approx1.
\end{equation}
Then, it follows from \eqref{prob_by_Dir_wigg}, \eqref{meas_trans}, \eqref{int_deN} that
\begin{equation}\label{energ_distr}
    \frac{dP(s,m,k_0)}{dk_0}\approx\pi N\al\frac{K^2}{\ga^2}\frac{q+q^{-1}}{2\omega} \Big(\frac{2}{m}\Big)^{4/3}\Big[\Ai'(y) +sn_k\Big(1-q\frac{1+n_k^2}{2n_k^2}\Big)\Big(\frac{m}{2}\Big)^{1/3}\Ai(y) \Big]^2,
\end{equation}
where
\begin{equation}
    y\approx \Big(\frac{m}{2}\Big)^{2/3} \Big[1-\frac{4n_k^2}{(1+n_k^2)^2q^2}\Big],
\end{equation}
and $n_k$ should be expressed in terms of $k_0^m$ from \eqref{spectrum_sol_app}, \eqref{n_k_def}. The Airy function and its derivative drop exponentially to zero for $y\gtrsim1/2$. Therefore, the radiation of twisted photons is exponentially suppressed when
\begin{equation}\label{m_max1}
    m\gtrsim m_c,\qquad m_c:=\min((1-q^{-2})^{-3/2},K^3)/\sqrt{2},
\end{equation}
The second quantity in the $\min$ function appearing in the definition of $m_c$ comes from the estimate of subsequent terms of the expansion of $y$ in $K^{-1}$ (see [(120), \cite{BKL2}]).

The distribution \eqref{energ_distr} reaches the maximum at
\begin{equation}\label{n_k_max}
    n_k\approx q-s c_0(2/m)^{1/3},
\end{equation}
where $c_0\approx 0.58$ is found from the equation
\begin{equation}
    (\Ai'(c^2_0)-c_0\Ai(c_0^2))'=0.
\end{equation}
If $m< m_c$, then
\begin{equation}\label{dPko}
    \frac{dP(m,k_0)}{dk_0}=\sum_{s=\pm1}\frac{dP(s,m,k_0)}{dk_0}
\end{equation}
possesses the maxima at points \eqref{n_k_max} with $s=\pm1$ and the local minimum at the point $n_k\approx q$. For $m\gtrsim m_c$, the maxima coalesce in the point $n_k\approx q$.

Such a behavior of radiation maxima is expectable, if one bears in mind that the wiggler radiation is just the synchrotron one in the Lorentz frame where the electron is at rest on average. The intensity profiles of lower synchrotron harmonics, $m\ll m_c$, were thoroughly investigated in Sec. 1.3.4 of \cite{Bord.1} where the effect of a blossoming out rose was revealed. The maxima of intensity of these harmonics do not lie in the orbit plane in the ultrarelativistic limit. In the orbit plane, the intensity of these harmonics possesses a local minimum. For large harmonics, $m\gtrsim m_c$, this minimum disappears. The maxima and minima of intensities of  synchrotron harmonics in the laboratory frame are found from the standard transformation law for angles
\begin{equation}\label{Lorentz_ang}
    \sin\theta=\frac{\sin\theta'}{1+[1-(1+K^2)/\gamma^2]^{1/2}\cos\theta'}\frac{\sqrt{1+K^2}}{\ga},
\end{equation}
where $\theta'$ is the polar angle counted in the ``synchrotron'' frame and $\theta$ is the same angle in the laboratory frame. In the synchrotron frame, the electron has the Lorentz factor
\begin{equation}
    \ga_s=\sqrt{1+K^2},
\end{equation}
and is ultrarelativistic in the wiggler case. The orbit plane, $\theta'=\pi/2$, is seen in the laboratory frame at the angle $\theta=\arcsin(\sqrt{1+K^2}/\ga)\approx K/\ga$. We shall return to the effect of blossoming out rose in Sec. \ref{Laser}.

Now we find a loose estimate for the number of twisted photons radiated by the right-handed helical wiggler at the harmonic $n=m$. The function \eqref{energ_distr} is peaked at $n_k\approx1$. The width of this peak can be found from the equation
\begin{equation}
    1-\frac{4n_k^2}{(1+n_k^2)^2q^2}=b\Big(\frac{2}{m}\Big)^{2/3},
\end{equation}
where $b\sim1$. Solving this equation, we obtain
\begin{equation}
    \De n_k\approx 2q \big[b(2/m)^{2/3}+q^{-2}-1\big]^{1/2}.
\end{equation}
Take into account that
\begin{equation}
    dk_0=-\frac{\ups_3}{n_3}\frac{k_0^2}{\omega m}\frac{K^2}{\ga^2} n_kdn_k.
\end{equation}
Then, assuming $(q-1)m^{1/3}\ll1$ and multiplying the value of $dP/d n_k$ at the point $n_k=1$ by $\De n_k$, we arrive at
\begin{equation}
    \De P(s,m)\approx 4\pi\al N(2/m)^{1/3}\big[b(2/m)^{2/3}+q^{-2}-1\big]^{1/2}\Ai'^2\Big(\Big(\frac{m}{2}\Big)^{2/3}(1-q^{-2})\Big).
\end{equation}
This quantity is independent of $K$. In the classical limit, $q=1$, we have
\begin{equation}\label{phot_raded}
    \De P(s,m)\approx 1.1\times 10^{-2} Nm^{-2/3},
\end{equation}
for $b=1.3$ (in [(135), \cite{BKL2}], the other quantity was estimated). Such a value of $b$ is taken for concordance of the estimate with the numerical calculations. This estimate shows, in particular, that, in describing the leading contribution to radiation of twisted photons produced in wigglers, the one-photon approximation is justified when
\begin{equation}\label{1photon}
    Nm^{-2/3}\lesssim 10.
\end{equation}
If this condition is violated, the trajectory can be partitioned into pieces such that condition \eqref{1photon} is satisfied for each part of the trajectory. The probabilities of radiation from different parts of the trajectory should be summed with account for a change of the electron energy-momentum due to radiation reaction on each part of the trajectory. In the classical regime, $q-1\ll1$, the Landau-Lifshitz equation can be employed to describe the effective electron dynamics \cite{PiMuHaKermp.2,BSBM18,Cole18,Poder18,LandLifshCTF.2,JacksonCE,Baryshev,KrivTsyt,SchlTikh.2,KazShiplde,KazAnn,BogKaz,RuKhaRyk,Neil18,HuChRu}.

\paragraph{Applicability conditions.}

Let us find the domain of applicability of the above formulas for the radiation of twisted photons in undulators with the quantum recoil taken into account. To shorten formulas, we suppose that all the dimensional quantities are measured in the units of the electron rest energy or in the electron Compton wavelengths.

Formula \eqref{m_max1} implies that the quantum recoil diminishes the maximum attainable value of $m$ for the twisted photons generated in the forward undulator radiation. Another restriction on the maximum $m$ follows from the requirement \eqref{trans_coh}. If
\begin{equation}\label{dipole_twist}
    n_\perp\ga\s_\perp\ll1,
\end{equation}
then \eqref{trans_coh} holds. The condition \eqref{dipole_twist} can be satisfied only in the dipole regime. In the wiggler case, the main part of radiation is produced with $n_\perp\ga\approx K\gtrsim3$ and even for the wave packet waist \eqref{sigma_perp_m} estimate \eqref{dipole_twist} is not fulfilled.

If
\begin{equation}
    n_\perp\ga\s_\perp \gtrsim1,
\end{equation}
then \eqref{trans_coh}, \eqref{spectrum_sol_app} imply
\begin{equation}
    \bar{k}_0^m/\ga=q-1\lesssim1/(10 n_\perp\ga\s_\perp)\;\Rightarrow\;q-1\ll1,
\end{equation}
i.e., in radiating a photon, the quantum recoil experienced by the electron should be small. In the wiggler case, for $n_k\approx1$, we obtain
\begin{equation}\label{m_max21}
    m\lesssim K/(10\omega\ga\s_\perp)= R/(10\s_\perp),
\end{equation}
where $K=\omega\ga R$ and $R$ is the radius of the spiral turn along which the electron is moving. Taking into account \eqref{sigma_perp_m}, we deduce the upper estimate
\begin{equation}\label{m_max22}
    m\lesssim RK/10,
\end{equation}
where recall that $R$ is measured in the Compton wavelengths $\lambdabar_C$. The estimates \eqref{m_max21}, \eqref{m_max22} are the necessary condition for the approximation of a point particle can be used. In this case, the localized wave packet of a particle radiating twisted photons in a wiggler can be characterized only by the average coordinate and momentum. As we see, the quantum recoil should be small for that to be the case.

It is useful to write restrictions \eqref{m_max1}, \eqref{m_max21} as a system of inequalities specifying the admissible region on the plane $(k_0^m/\e,m)$:
\begin{equation}\label{admis_reg}
    \frac{k_0^m}{\ga}\lesssim\frac14\Big(\frac{2}{m}\Big)^{2/3},\qquad m\lesssim\frac{K^3}{\sqrt{2}},\qquad \frac{k_0^m}{\ga}\lesssim\frac1{10 K\s_\perp}.
\end{equation}
Note that $k_0^m\approx\bar{k}_0^m$. One can distinguish two cases
\begin{equation}
    i)\;5\s_\perp>K,\qquad ii)\;5\s_\perp<K.
\end{equation}
In the case (i), the region \eqref{admis_reg} is reduced to a rectangle
\begin{equation}\label{case_i}
     \frac{k_0^m}{\ga}\lesssim\frac1{10 K\s_\perp},\qquad m\lesssim\frac{K^3}{\sqrt{2}}.
\end{equation}
The twisted photons with the maximum energy and projection of the total angular momentum are produced when inequalities \eqref{case_i} turn into the equalities. In the case (ii), the region \eqref{admis_reg} has nontrivial angular points at
\begin{equation}
    k_0^m=\frac{\ga}{10K\s_\perp},\quad m=\frac{(5K\s_\perp)^{3/2}}{\sqrt{2}};\qquad k_0^m=\frac{\ga}{2K^2},\quad m=\frac{K^3}{\sqrt{2}},
\end{equation}
which coalesce for $5\s_\perp=K$. In terms of $m$, the system of inequalities \eqref{admis_reg} is written as
\begin{equation}\label{m_ineq}
    m\lesssim\frac12\Big(\frac{K^2}{\ga\omega}\Big)^{3/5},\qquad m\lesssim\frac{K^3}{\sqrt{2}},\qquad m\lesssim\frac{K}{10\s_\perp\ga\omega},
\end{equation}
respectively.

\section{Radiation by charged particles in the laser wave}\label{Laser}

Let us apply the above general formulas for description of radiation of twisted photons by an ultrarelativistic charged particle in the laser wave of a circular polarization. We suppose that the one-photon radiation gives the leading contribution to radiation of twisted photons. Then we can employ the formulas from the preceding sections to describe this radiation. The strength tensor of the electromagnetic field reads as follows (we use the notation borrowed from \cite{KazAnn,BogKaz,BKL1})
\begin{equation}\label{fmunu_ew.1}
    eF^{\mu\nu}=a(\xi)h_-^{[\mu}\left[h_1^{\nu]}\cos\vf(\xi)+h_2^{\nu]}\sin\vf(\xi)\right],
\end{equation}
where $h^\mu_-=(1,0,0,\zeta)$, $h^\mu_{1,2}=\de^\mu_{1,2}$, the function $a(\xi)$ characterizes the amplitude of the electromagnetic field and $\vf(\xi)$ is the phase, where $\xi=h^\mu_-x_\mu=x^0-\zeta x^3$. We consider the situation when the electromagnetic wave propagates along the axis of the detector of twisted photons. The quantity $\zeta=\pm1$, where the upper sign corresponds to the case when the wave moves towards the detector and the lower sign is for the case when the wave moves from the detector.

It is useful to convert all the quantities to the dimensionless ones using the Compton wavelength as a unit length [(5), \cite{BKL1}]. Then, for example, the laser wave with intensity $10^{22}$ W/cm${}^2$ and photon energy $1.53$ eV \cite{laser_tod} corresponds to
\begin{equation}\label{las_wave}
    |a|\approx1.47\times 10^{-4},\qquad|\Omega|\approx2.99\times10^{-6}.
\end{equation}
The Lorentz equations are easily solved with arbitrary function $a(\xi)$ (see, e.g., [(51), \cite{KazAnn}] for $\la=0$ and \cite{LandLifshCTF.2}):
\begin{equation}\label{Lorentz_sol}
\begin{split}
    x^1(\xi)&=x^1(0)+r^1(0)\xi -\ups_-^{-1}\int_0^\xi dx\int_0^x dy a(y)\cos\vf(y),\\
    x^2(\xi)&=x^2(0)+r^2(0)\xi -\ups_-^{-1}\int_0^\xi dx\int_0^x dy a(y)\sin\vf(y),\\
    x^0(\xi)+\zeta x^3(\xi)&=\int_0^\xi dx\Big[\ups_-^{-2} +\Big(r^1(0)-\ups_-^{-1}\int_0^x dya(y)\cos\vf(y)\Big)^2+\\ &+\Big(r^2(0)-\ups_-^{-1}\int_0^x dya(y)\sin\vf(y)\Big)^2\Big],\\
    \xi&=\ups_-\tau,
\end{split}
\end{equation}
where $r^\mu:=\ups^{\mu}/\ups_-$, $\ups^\mu$ is the $4$-velocity, $\ups_\mu\ups^\mu=1$, $\ups_-:=\ups^0-\zeta\ups^3=const$, $\tau$ is the proper time, and it is assumed that $x^0=x^3=0$ at the initial instant of time. It is clear this assumption does not destroy the generality of our considerations. Upon shifting $x^0$ and $x^3$ by constants, the amplitude of radiation of a twisted photon changes by an overall phase, which does not affect the probability to record the twisted photon by a detector. It is convenient to pass in formulas \eqref{prob_by_scal}, \eqref{prob_by_Dir} and [(36), \cite{BKL2}] from the integration variable $t$ to the variable $\xi$. The corresponding derivatives take the form
\begin{equation}\label{Lorentz_sol_1}
\begin{split}
    r^1(\xi)&=r^1(0) -\ups_-^{-1}\int_0^\xi dx a(x)\cos\vf(x),\\
    r^2(\xi)&=r^2(0) -\ups_-^{-1}\int_0^\xi dx a(x)\sin\vf(x),\\
    r^0(\xi)+\zeta r^3(\xi)&=\ups_-^{-2} +\Big(r^1(0)-\ups_-^{-1}\int_0^\xi dxa(x)\cos\vf(x)\Big)^2 +\Big(r^2(0)-\ups_-^{-1}\int_0^\xi dxa(x)\sin\vf(x)\Big)^2.
\end{split}
\end{equation}

In order to obtain analytic formulas, we assume that the phase
\begin{equation}
    \vf=\Omega \xi+\vf_0,
\end{equation}
where $\Omega$ is the frequency of the electromagnetic wave and $\vf_0$ is the initial phase. The amplitude is chosen as
\begin{equation}\label{envelope_const}
    a(\xi)=const,\quad\xi\in[0,2\pi N],
\end{equation}
where $N$ is the number of periods of the electromagnetic wave. The amplitude vanishes outside this interval, i.e., it is assumed that the laser wave pulse possesses sharp rising and descending edges. In that case, employing the notation from \eqref{prob_by_scal}, \eqref{prob_by_Dir}, and [(36), \cite{BKL2}], we have
\begin{equation}\label{r_plane_wave}
\begin{gathered}
    r_\pm=\bar{r}_\pm \pm iK\ups_-^{-1}e^{\pm i\vf},\qquad r^0=\frac1{2\ups_-^2}\big[\ups_-^2+1+K^2+\bar{\ups}_\perp^2 -2K\bar{\ups}_\perp\sin(\vf-\rho) \big],\\
    r^3=\frac{\zeta}{2\ups_-^2}\big[1+K^2+\bar{\ups}_\perp^2 -\ups_-^2 -2K\bar{\ups}_\perp\sin(\vf-\rho) \big],
\end{gathered}
\end{equation}
where $K:=a/\Omega$,
\begin{equation}\label{r_plane_wave_1}
    \bar{r}_\pm:=r_\pm(0)\mp iK\ups_-^{-1}e^{\pm i\vf_0},\qquad\bar{\ups}_\perp:=\ups_-|\bar{r}_\pm|
\end{equation}
and $\rho=\arg\bar{r}_+$. The solution to the Lorentz equations is given by
\begin{equation}\label{x_sol}
\begin{gathered}
    x_\pm=\bar{x}_\pm +\bar{r}_\pm\xi+\frac{K}{\ups_-\Omega}e^{\pm i\vf},\qquad
    x^0=\frac1{2\ups_-^2}\big[(\ups_-^2+1+K^2+\bar{\ups}_\perp^2)\xi +2\bar{\ups}_\perp\frac{K}{\Omega} (\cos(\vf-\rho)-\cos(\vf_0-\rho)) \big],\\
    x^3=\frac{\zeta}{2\ups_-^2}\big[(1+K^2+\bar{\ups}_\perp^2-\ups_-^2)\xi +2\bar{\ups}_\perp\frac{K}{\Omega}(\cos(\vf-\rho)-\cos(\vf_0-\rho)) \big],
\end{gathered}
\end{equation}
where
\begin{equation}\label{x_bar}
    \bar{x}_\pm:=x_\pm(0)-\frac{K}{\ups_-\Omega}e^{\pm i\vf_0}.
\end{equation}
Notice that if the charged particle moves initially along the axis of the twisted photon detector, viz., $r_\pm(0)=0$, then $\bar{r}_\pm\neq0$.

\paragraph{Radiation without recoil.}

Let us consider the radiation of twisted photons in the case when the quantum recoil is negligible. In order to find the probability of radiation [(36), \cite{BKL2}], it is necessary to evaluate the amplitudes
\begin{equation}\label{I_3pm}
\begin{split}
    I_3&=\int_0^{TN}d\xi r^3(\xi)e^{-ik_0(x^0-n_3x^3)}j_m(k_\perp x_+,k_\perp x_-),\\
    I_\pm&=\frac{in_\perp}{s\mp n_3}\int_0^{TN}d\xi r_\pm(\xi)e^{-ik_0(x^0-n_3x^3)}j_{m\mp1}(k_\perp x_+,k_\perp x_-).
\end{split}
\end{equation}
Then the probability to record the twisted photon is
\begin{equation}
    dP(s,m,k_\perp,k_3)=e^2\Big|I_3+\frac12I_++\frac12I_-\Big|^2n_\perp^3\frac{dk_3dk_\perp}{16\pi^2}.
\end{equation}
In formula \eqref{I_3pm}, the contributions from the parts of particle trajectory with $\xi\not\in[0,2\pi N]$ are neglected. These contributions correspond to the edge radiation. They can be ignored when the energies of recorded photons are sufficiently large. A thorough description of the edge radiation in terms of twisted photons is given in \cite{BKL3}.

The evaluation of integrals \eqref{I_3pm} is performed analogously to the case of undulator radiation studied in \cite{BKL2}. First of all, we shift the integration variable $\xi\rightarrow\xi+TN/2$ and make use of the addition theorem [(A6), \cite{BKL2}] for the Bessel functions (see also \cite{Wats.6}):
\begin{equation}\label{add_thm}
    j_m(k_\perp x_+,k_\perp x_-)=\sum_{l=-\infty}^\infty j_{m-l}\big(k_\perp z_+,k_\perp z_-\big) j_l(k_\perp y_+,k_\perp y_-),
\end{equation}
where
\begin{equation}\label{zpm}
    z_\pm:=\bar{x}_\pm +\bar{r}_\pm TN/2,\qquad y_\pm:=\bar{r}_\pm\xi+\frac{K}{\ups_-\Omega}e^{\pm i\vf}.
\end{equation}
Now the phase $\vf_0$ entering into $\vf$ includes $\pi N$. We denote this phase by $\vf_{0N}$. Notice that, on shifting the variable $\xi$, the phase $\vf_0$ appearing explicitly in formulas \eqref{r_plane_wave_1}, \eqref{x_sol}, and \eqref{x_bar} does not change. Substitute the integral representation [(A8), \cite{BKL2}]
\begin{equation}
    j_l(k_\perp y_+,k_\perp y_-)=i^{-l}\int_{-\pi}^\pi\frac{d\psi}{2\pi} e^{il\psi}e^{ik_\perp(y_2\sin\psi+y_1\cos\psi)}
\end{equation}
into \eqref{add_thm} and then \eqref{add_thm} into \eqref{I_3pm}. As a result, the expression standing in the exponent in the integrand of $I_3$ becomes
\begin{equation}
    -i\frac{k_0}{2\ups_-^2}\xi\big[(1-\zeta n_3)(1+K^2+\bar{\ups}_\perp^2)+(1+\zeta n_3)\ups_-^2 -2n_\perp\bar{\ups}_\perp\ups_-\cos(\psi-\rho) \big]+i\eta\sin(\vf+\de)  +il\psi,
\end{equation}
up to a constant term that does not influence the probability of radiation. Here
\begin{equation}
    \eta\cos\de:=\frac{Kk_0}{\ups_-\Omega}(n_\perp\sin\psi -(1-\zeta n_3)\bar{r}^2),\qquad \eta\sin\de:=\frac{Kk_0}{\ups_-\Omega}(n_\perp\cos\psi -(1-\zeta n_3)\bar{r}^1).
\end{equation}
Using the Fourier series expansion
\begin{equation}\label{Bessel_generate}
    e^{i\eta\sin(\vf+\de)}=\sum_{n=-\infty}^\infty e^{in(\vf+\de)}J_n(\eta),
\end{equation}
the integral over $\xi$ is reduced to the delta-like sequence
\begin{equation}
    \int_{-TN/2}^{TN/2}\frac{d\xi}{2\pi}e^{-ix_n\xi}=\de_N(x),\qquad\de_N(x_n):=\frac{\sin(TNx_n/2)}{\pi x_n}.
\end{equation}
The argument of the regularized delta function reads
\begin{equation}
    x_n=\frac{k_0}{2\ups_-^2}\big[(1-\zeta n_3)(1+K^2+\bar{\ups}_\perp^2)+(1+\zeta n_3)\ups_-^2 -2n_\perp\bar{\ups}_\perp\ups_-\cos(\psi-\rho) \big]-\Omega n.
\end{equation}
For $N$ large, the main contribution to the integral over $\psi$ comes from the points where the argument of the regularized delta function vanishes.

Below we shall assume that $\Omega>0$ and, at the end, shall discuss how the results change for $\Omega<0$. The condition $x_n=0$ can be conveniently written as
\begin{equation}\label{delta_arg}
    x_n=\frac{\Omega}{2}(b_n+a_n)[\cos\xi_n-\cos(\psi-\rho)]=0,
\end{equation}
where the notation has been introduced \cite{BKL2}
\begin{equation}\label{a_n_b_n}
    a_n:=n-k_0\omega_+^{-1},\qquad b_n:=k_0\omega_-^{-1}-n,\qquad \xi_n:=\arccos\frac{b_n-a_n}{b_n+a_n},
\end{equation}
and
\begin{equation}\label{omega_pm}
    \omega_\pm:=\frac{2\Omega\ups_-^2}{(1-\zeta n_3)(1+K^2+\bar{\ups}_\perp^2)+(1+\zeta n_3)\ups_-^2\mp 2n_\perp\bar{\ups}_\perp\ups_-}.
\end{equation}
If $N$ is so large that $\de_N(x_n)$ removes the integration over $\psi$, then the energy spectrum of radiated twisted photons consists of the intervals
\begin{equation}\label{spectrum_laser}
    k_0\in n[\omega_-,\omega_+],\quad n=\overline{1,\infty}.
\end{equation}
The radiation is suppressed outside these intervals. These intervals become overlapping starting from the harmonic number
\begin{equation}\label{overlap}
    n_0=\frac{\omega_-}{\omega_+-\omega_-}.
\end{equation}
When $k_0$ belongs to the intervals, $a_n\geqslant0$, $b_n\geqslant0$, and $\xi_n\in[0,\pi]$.

As a result, neglecting the terms at nonpositive $n$, we have
\begin{multline}\label{I_3_las}
    I_3\approx \sum_{n=1}^\infty \sum_{l=-\infty}^\infty j_{m-l}(k_\perp z_+,k_\perp z_-)\int_{-\pi}^\pi d\psi \de_N(x_n)  e^{in(\de+\vf_{0N})+il\psi}i^{-l} \times\\
     \times\frac{\zeta}{2\ups_-^2}\Big[ (1+K^2+\bar{\ups}_\perp^2 -\ups_-^2)J_n(\eta) +iK\bar{\ups}_\perp \big(e^{-i(\de+\rho)}J_{n-1}(\eta) -e^{i(\de+\rho)}J_{n+1}(\eta)\big) \Big].
\end{multline}
As for the rest integrals, we obtain similarly
\begin{equation}\label{I_pm}
    I_\pm\approx \sum_{n=1}^\infty \sum_{l=-\infty}^\infty j_{m-l}(k_\perp z_+,k_\perp z_-)\int_{-\pi}^\pi d\psi \de_N(x_n)  e^{in(\de+\vf_{0N})+i(l\mp1)\psi}i^{-l}
    \frac{\mp n_\perp}{s\mp n_3}\Big[\bar{r}_\pm J_n \pm iK\ups_-^{-1}J_{n\mp1}e^{\mp i\de} \Big],
\end{equation}
where the arguments of the Bessel functions are the same as in \eqref{I_3_las}. In order to obtain \eqref{I_pm}, one needs to shift the summation index $l\rightarrow l\mp1$ in the series \eqref{add_thm}. Taking into account estimates \eqref{estimates}, the total contribution to the radiation amplitude takes the form
\begin{equation}\label{tot_ampl_laser}
    I_3+\frac12(I_++I_-)\approx\sum_{n=1}^\infty \sum_{l=-\infty}^\infty j_{m-l}(k_\perp z_+,k_\perp z_-)\int_{-\pi}^\pi \frac{d\psi}{2} \de_N(x_n)  e^{in(\de+\vf_{0N})+il\psi}i^{-l}
    g_n(\psi),
\end{equation}
where
\begin{equation}
\begin{split}
    g_n(\psi):=&\,\zeta\frac{1+K^2+\bar{\ups}_\perp^2-\ups_-^2}{\ups_-^2}J_n
    +\zeta \frac{iK\bar{\ups}_\perp}{\ups_-^2} \big(e^{-i(\de+\rho)}J_{n-1} -e^{i(\de+\rho)}J_{n+1}\big)-\\
    &-\frac{2}{n_\perp}e^{-is\psi}\big(\bar{r}_s J_n+s \frac{i K}{\ups_-} J_{n-s}e^{-is\de}\big).
\end{split}
\end{equation}
The last expression can be rewritten in terms of the Bessel function and its derivative with the same index
\begin{equation}
\begin{split}
    g_n(\psi)=&\,2\Big[\zeta\frac{1+K^2+\bar{\ups}_\perp^2-\ups_-^2}{2\ups_-^2} -\frac{isK n}{n_\perp\ups_-\eta}e^{-is(\de+\psi)} +\zeta\frac{K\bar{\ups}_\perp n}{\ups_-^2\eta}\sin(\de+\rho) -\frac{\bar{r}_s}{n_\perp}e^{-is\psi} \Big]J_n-\\
    &-\frac{2iK}{n_\perp\ups_-}\Big[e^{-is(\de+\psi)} -\zeta\frac{n_\perp\bar{\ups}_\perp}{\ups_-}\cos(\de+\rho) \Big]J'_n.
\end{split}
\end{equation}
Further, we suppose that $N$ is so large that $\de_N(x_n)$ removes the integration over $\psi$. The solution of \eqref{delta_arg} is, evidently,
\begin{equation}
    \psi=\rho\pm\xi_n.
\end{equation}
As in the case of undulator radiation, the three cases occur \cite{BKL2}: (a) the regular case $\xi_n\neq\{0,\pi\}$; (b) the weakly degenerate case $\xi_n=\{0,\pi\}$; and (c) the strongly degenerate case $a_n=b_n=0$.

Let us begin with the regular case. For $N$ large, in the leading order, we deduce
\begin{equation}\label{deltaN}
    \de_N(x_n)\approx\frac{\theta(a_n)\theta(b_n)}{\Omega\sqrt{a_nb_n}}[\de(\psi-\rho-\xi_n)+\de(\psi-\rho+\xi_n)].
\end{equation}
The delta functions remove integration in \eqref{tot_ampl_laser}. The remaining sum over $l$ can be performed by using the relation
\begin{equation}
    \sum_{k=-\infty}^\infty t^k j_k(p,q)=e^{(pt-q/t)/2},
\end{equation}
which follows from [(A7), \cite{BKL2}]. Then, up to an irrelevant phase,
\begin{equation}
\begin{split}
    I_3+\frac12(I_++I_-)\approx\sum_{n=1}^\infty \frac{\theta(a_n)\theta(b_n)}{2\Omega\sqrt{a_nb_n}}e^{in\vf_{0N}} &\Big\{e^{in\de+im\psi+i k_\perp|z_+|\cos(\psi-\arg z_+)}g_n(\psi)\Big|_{\psi=\rho+\xi_n}+\\ &+e^{in\de+im\psi +i k_\perp|z_+|\cos(\psi-\arg z_+)}g_n(\psi)\Big|_{\psi=\rho-\xi_n} \Big\}.
\end{split}
\end{equation}
In the photon energy range where the harmonics do not overlap, we obtain
\begin{equation}\label{regul_laser}
\begin{split}
    dP(s,m,k_\perp,k_3)\approx e^2\sum_{n=1}^\infty \frac{\theta(a_n)\theta(b_n)}{4\Omega^2a_nb_n}&\bigg|e^{in\de+im\psi +i k_\perp|z_+|\cos(\psi-\arg z_+)}g_n(\psi)\Big|_{\psi=\rho+\xi_n}+\\
    &+e^{in\de+im\psi +i k_\perp|z_+|\cos(\psi-\arg z_+)}g_n(\psi)\Big|_{\psi=\rho-\xi_n} \bigg|^2  n_\perp^3\frac{dk_3dk_\perp}{16\pi^2}.
\end{split}
\end{equation}
The dependence on $m$ for the energies $k_0$ belonging to the spectral band with number $n$ is periodic with the period \cite{BKL2}
\begin{equation}\label{Tm_period}
    T_m=\left\{
         \begin{array}{ll}
           \pi/\xi_n, & \xi_n\in(0,\pi/2); \\
           \pi/(\pi-\xi_n), & \xi_n\in[\pi/2,\pi).
         \end{array}
       \right.
\end{equation}
Of course, this periodicity holds only for those quantum numbers $m$ where $\de_N(x_n)$ can be replaced by delta function \eqref{deltaN}.

Now we turn to the weakly degenerate case. Let $a_n=0$, $b_n>0$, i.e., $k_0=n\omega_+$. Then
\begin{equation}
    b_n=n(\omega_+\omega_-^{-1}-1),\qquad\xi_n=0,\qquad\psi=\rho.
\end{equation}
We assume that $N$ is so large that all the integrand functions in \eqref{tot_ampl_laser}, apart form $\de_N(x_n)$, can be taken at the point $\psi=\rho$ and be removed from the integrand. In that case, the integral arises
\begin{equation}
    \int_{-\pi}^\pi d\psi\de_N(x_n)\approx\int_{-\infty}^\infty d\psi\frac{\sin\big[\pi Nn(\omega_+\omega_-^{-1}-1)\psi^2/4\big]}{\pi\Omega n(\omega_+\omega_-^{-1}-1)\psi^2/4}=\Omega^{-1}\sqrt{\frac{8N}{n(\omega_+\omega_-^{-1}-1)}}.
\end{equation}
The probability to record a twisted photon becomes
\begin{equation}\label{deg_laser1}
    dP(s,m,k_\perp,k_3)\approx e^2 \frac{N |g_n(\rho)|^2n_\perp^3}{\Omega^2 n(\omega_+\omega^{-1}_--1)}  \frac{dk_3dk_\perp}{8\pi^2}.
\end{equation}
For $b_n=0$, $a_n>0$, i.e., for $k_0=n\omega_-$, the similar calculations lead to
\begin{equation}\label{deg_laser2}
    dP(s,m,k_\perp,k_3)\approx e^2 \frac{N |g_n(\pi-\rho)|^2n_\perp^3}{\Omega^2 n(1-\omega_-\omega^{-1}_+)}  \frac{dk_3dk_\perp}{8\pi^2}.
\end{equation}
In the domain where the applicability conditions of the approximations made are fulfilled, the explicit dependence of the twisted photon radiation probability on $m$ disappears.

In the strongly degenerate case, the spectral bands \eqref{spectrum_laser} turn into narrow lines $k_0=n\omega_+=n\omega_-$. This happens when $\bar{\ups}_\perp\approx0$. Then
\begin{equation}\label{forw_chi_psi}
    \eta=\frac{Kk_\perp}{\ups_-\Omega},\qquad \de=\pi/2-\psi,
\end{equation}
The functions,
\begin{equation}\label{xn_gn_forw_las}
\begin{split}
    x_n&=\frac{k_0}{2\ups_-^2}\big[(1-\zeta n_3)(1+K^2)+(1+\zeta n_3)\ups_-^2 \big]-\Omega n,\\ g_n&=\zeta\frac{1+K^2-\ups_-^2}{\ups_-^2}J_n-\frac{2K}{n_\perp \ups_-} J_{n-s}=\Big[\zeta\frac{1+K^2-\ups_-^2}{\ups_-^2}-\frac{2\Omega n}{n_\perp k_\perp}\Big]J_n -\frac{2sK}{n_\perp\ups_-}J'_n,
\end{split}
\end{equation}
do not depend on $\psi$, and the integral over $\psi$ in \eqref{tot_ampl_laser} is readily performed. As a result,
\begin{equation}
    I_3+\frac12(I_++I_-)\approx \pi\de_N(x_n) j_{m-n}(k_\perp z_+,k_\perp z_-)e^{in\vf_0}(-1)^{nN}g_n.
\end{equation}
The probability to record a twisted photon is given by
\begin{equation}\label{forw_laser}
    dP(s,m,k_\perp,k_3)=e^2 \de_N^2(x_n) J_{m-n}^2(k_\perp |z_+|)g_n^2 n_\perp^3\frac{dk_3dk_\perp}{16}.
\end{equation}
If $k_\perp|z_+|\ll1$, then the selection rule $m=n$ is fulfilled \cite{KatohPRL,TaHaKa,BKL2,SasMcNu,HMRR,HeMaRo,BHKMSS,HKDXMHR,KatohSRexp,Rubic17}.

Let us show how the above results are modified for $\Omega<0$. The sign change of $\Omega$ corresponds to a change of polarization of the incident electromagnetic wave \eqref{fmunu_ew.1}. This, in turn, leads to a change of  handedness of the helix along which the charged particle is moving. Upon changing the sign of $\Omega$, formulas \eqref{a_n_b_n}-\eqref{overlap} remain valid with the replacement $\Omega\rightarrow|\Omega|$. Since the substitution $\Omega\rightarrow-\Omega$ results in
\begin{equation}
    K\rightarrow-K,\qquad n\rightarrow-n,
\end{equation}
we have
\begin{equation}
    g_n(\psi)\rightarrow(-1)^ng_n(\psi).
\end{equation}
Therefore, on substituting $\Omega\rightarrow-|\Omega|$ in \eqref{regul_laser}, \eqref{deg_laser1}, \eqref{deg_laser2}, and \eqref{forw_laser}, one must set
\begin{equation}
    e^{in\de}\rightarrow e^{-in\de},
\end{equation}
in formula \eqref{regul_laser}, formulas \eqref{deg_laser1}, \eqref{deg_laser2} remain intact, and, in formula \eqref{forw_laser}, one needs to replace
\begin{equation}
    J_{m-n}^2(k_\perp |z_+|)\rightarrow J_{m+n}^2(k_\perp |z_+|).
\end{equation}
For $k_\perp|z_+|\ll1$, the selection rule in the strongly degenerate case looks as $m=-n$.

Consider in more detail the cases when the electromagnetic wave propagates towards the detector of twisted photons or from it. In these cases, the electron bunch moves approximately along the direction of propagation of the electromagnetic wave or in the opposite direction, respectively. In the case when the electromagnetic wave propagates toward the detector, we have $\zeta=1$ and
\begin{equation}
    \ups_-\approx \frac{1+\ups_\perp^2}{2\ga}\sim\frac{\vk^2}{2\ga},\qquad \ups_\perp\sim\vk,\qquad \bar{\ups}_\perp\lesssim \vk,\qquad n_\perp\lesssim\frac{\vk}{\ga}.
\end{equation}
The bounds of the spectral bands \eqref{spectrum_laser} are expressed through
\begin{equation}\label{spectrum_1}
    \omega_\pm\approx \Omega\Big[\big(1\mp \frac{n_\perp\bar{\ups}_\perp}{2\ups_-}\big)^2+\frac{n_\perp^2}{4\ups_-^2}(1+K^2) \Big]^{-1}\sim\Omega.
\end{equation}
In the strongly degenerate case, the radiation spectrum becomes
\begin{equation}\label{spectrum_forw_1}
    k_0\approx \Omega n\Big[1+\frac{n_\perp^2}{4\ups_-^2}(1+K^2)\Big]^{-1},\qquad n=\overline{1,\infty}.
\end{equation}
For $n\geqslant5$, the Bessel functions entering into $g_n$ can be expressed through the Airy functions (see [(122), \cite{BKL2}] and also \cite{Bord.1}) with
\begin{equation}
    x=1-\frac{16K^2\ups_-^2/n_\perp^4}{\big(1+K^2+4\ups_-^2/n_\perp^2\big)^2}.
\end{equation}
For the radiation probability not to be exponentially suppressed, this quantity should be small $x\lesssim1/20$. This occurs for $K\gtrsim3$ and
\begin{equation}\label{n_perp_opt_1}
    n_\perp\approx\frac{2\ups_-}{K}\approx \frac{1+\ups_\perp^2}{\ga K}\sim\frac{\vk}{\ga}.
\end{equation}
In that case, the radiation probability drops exponentially to zero at the harmonic numbers \cite{BKL2}
\begin{equation}
    n\gtrsim\frac{K^3}{\sqrt{2}}.
\end{equation}

If the electromagnetic wave moves from the detector, i.e., the head-on collision of the laser wave with the bunch of charged particles is considered, then $\zeta=-1$ and
\begin{equation}\label{estim_headon}
    \ups_-\approx 2\ga,\qquad\bar{\ups}_\perp\lesssim\vk,\qquad n_\perp\lesssim\vk/\ga,
\end{equation}
and also
\begin{equation}\label{omega_pm_headon}
    \omega_\pm\approx\frac{\Omega\ups_-^2}{1+K^2+(\bar{\ups}_\perp\mp n_\perp\ups_-/2)^2}\sim\frac{4\Omega\ga^2}{\vk^2}.
\end{equation}
The analysis in this case is completely analogous to the analysis of the undulator radiation \cite{BKL2} at the observation angle $\theta:=\bar{\ups}_\perp/\ga$ and the undulator frequency $\omega:=2\Omega$ (cf. [(85), \cite{BKL2}]). In the strongly degenerate case, the radiation spectrum looks as
\begin{equation}\label{spectrum_forw_-1}
    k_0\approx\frac{\Omega n\ups_-^2}{1+K^2+n^2_\perp\ups_-^2/4},\qquad n=\overline{1,\infty}.
\end{equation}
For $n\geqslant5$, the radiation probability is not exponentially suppressed at $K\gtrsim3$ if $n_\perp\approx K/\ga$ and $n\lesssim K^3/\sqrt{2}$.

\paragraph{Radiation with recoil.}

Now we take the quantum recoil into account. In the case $\zeta=1$, estimate \eqref{spectrum_1} holds. Therefore, the quantum recoil can be neglected for reasonable photon energies of the laser wave (see \eqref{las_wave}). In that case, the probability to record a twisted photon radiated by both scalar and Dirac particles is described by formula [(36), \cite{BKL2}]. Thus, formulas obtained above remain intact with good accuracy.

In the case $\zeta=-1$, the quantum recoil can be significant. Since the approximate equality \eqref{estim_headon} is valid, the quantity
\begin{equation}\label{q_appr}
    q=P_0/P'_0=\ups_0/\ups'_0=\ga/(\ga-k_0)\approx \ups_-/(\ups_--2k_0)=const,
\end{equation}
up to the terms of order $\vk^2/\ga^2$. Hence, the probability to detect a twisted photon radiated by a charged scalar particle with the quantum recoil taken into account \eqref{prob_by_scal} is obtained from the formulas above, where the recoil was ignored, by multiplying the probability by $q$ and substituting
\begin{equation}\label{BK_rplmnt}
    k_0\rightarrow\bar{k}_0:=qk_0
\end{equation}
in the definitions of $a_n$, $b_n$ \eqref{a_n_b_n} and the radiation spectrum \eqref{spectrum_laser}. As was discussed in the preceding sections, the term standing in the exponent and proportional to $k_\perp^2/(2P_{0i})$ can be safely neglected. As far as the strongly degenerate case \eqref{forw_laser} is concerned, substitution \eqref{BK_rplmnt} has to be done in formula \eqref{xn_gn_forw_las} for $x_n$, and, of course, \eqref{forw_laser} must be multiplied by $q$.

The treatment of the Dirac particle case is a bit more complex. The probability to record a twisted photon equals
\begin{equation}\label{tot_prob}
    dP(s,m,k_3,k_\perp)=dP_1(s,m,k_\perp,k_3)+dP_a(s,m,k_\perp,k_3),
\end{equation}
where $dP_1(s,m,k_\perp,k_3)$ is the contribution of the first term in \eqref{prob_by_Dir} and $dP_a(s,m,k_\perp,k_3)$ is the contribution of the second and third terms in \eqref{prob_by_Dir}. As long as \eqref{q_appr} holds, the contribution of the first term in \eqref{prob_by_Dir} is evaluated as in the case of negligible quantum recoil: the probability to record a twisted photon without recoil must be multiplied by
\begin{equation}
    (1+q)^2/4,
\end{equation}
and substitution \eqref{BK_rplmnt} must be performed in the definitions of $a_n$, $b_n$ \eqref{a_n_b_n} and the radiation spectrum \eqref{spectrum_laser}.

The contribution of the last two terms in \eqref{prob_by_Dir} has to be evaluated from scratch. It follows from the explicit expressions for the mode functions [(13), \cite{BKL2}] that for $s=-1$ the third term in \eqref{prob_by_Dir} can be omitted while for $s=1$ the second term can be thrown out. Let
\begin{equation}
    I_a:=\frac{k_0}{4\ups_0'}\int_0^{TN} d\xi e^{-i\bar{k}_0(x^0-n_3x^3)}[r_sa_{-s}-isn_\perp r^0a_{-s}(m+s)].
\end{equation}
The contribution of the last two terms in \eqref{prob_by_Dir} is proportional to the modulus squared of this integral. For the head-on collision, we have
\begin{equation}
    r^0\approx1/2,
\end{equation}
Then, performing the calculations completely analogous to the case of negligible recoil, we find
\begin{equation}
    I_a\approx\frac{k_0}{4\ups_0'} \sum_{n=1}^\infty \sum_{l=-\infty}^\infty j_{m-l}(k_\perp z_+,k_\perp z_-)\int_{-\pi}^\pi d\psi \de_N(x_n)  e^{in(\de+\vf_{0N})+il\psi}i^{-l} g^a_n(\psi),
\end{equation}
where
\begin{equation}
    g^a_n(\psi):=\Big(1-2\frac{\bar{r}_s}{n_\perp}e^{-is\psi}\Big) J_n(\eta) -\frac{2isK}{n_\perp\ups_-}J_{n-s}(\eta)e^{-is(\de+\psi)},
\end{equation}
and $x_n$ has the form \eqref{delta_arg} with
\begin{equation}
    a_n:=n-\bar{k}_0\omega_+^{-1},\qquad b_n:=\bar{k}_0\omega_-^{-1}-n,\qquad \xi_n:=\arccos\frac{b_n-a_n}{b_n+a_n}.
\end{equation}
The approximate expressions for $\omega_\pm$ are written in \eqref{omega_pm_headon}.

\begin{figure}[t]
   \centering
\begin{tabular}{cc}
\includegraphics*[align=c,width=0.47\linewidth]{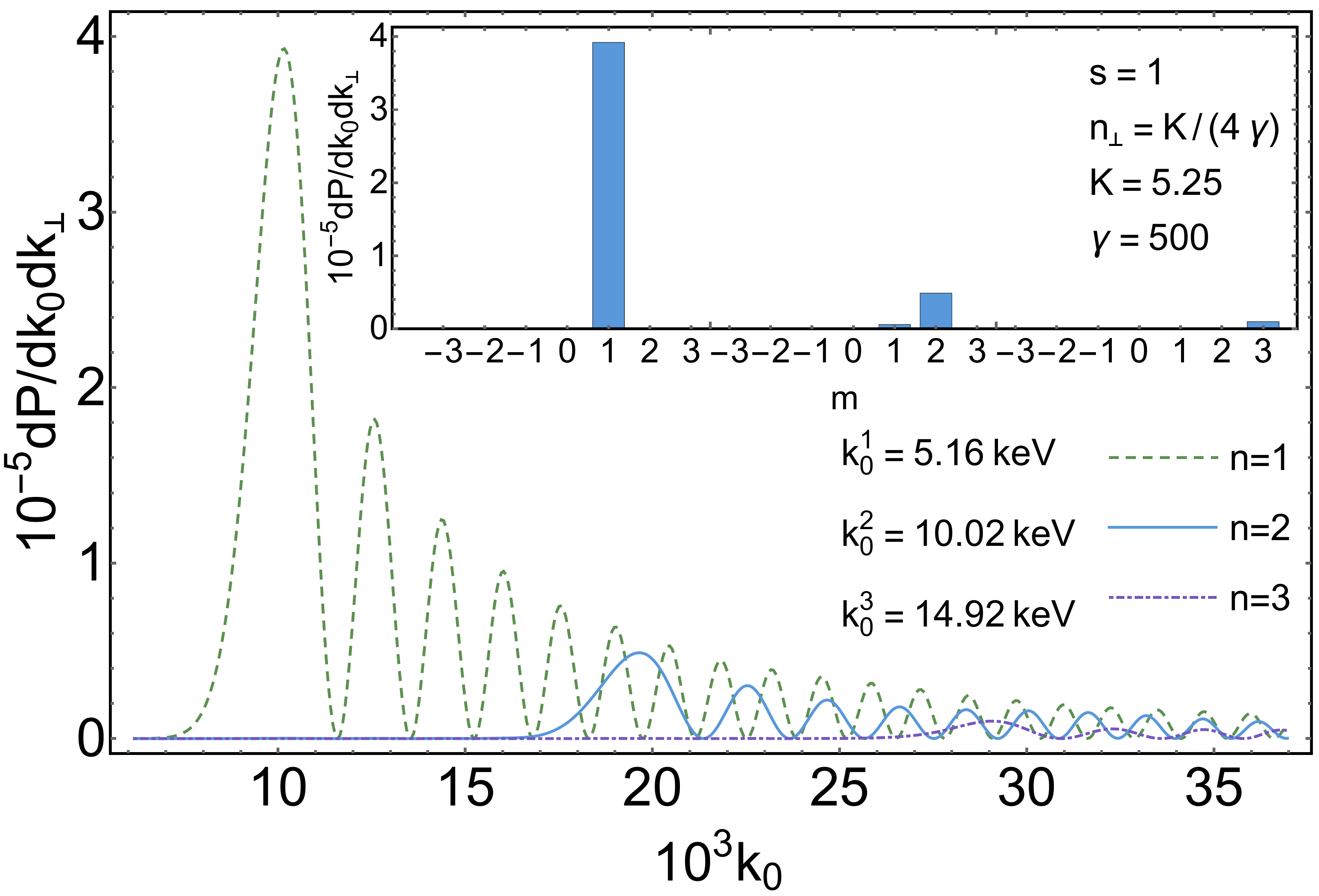}&
\includegraphics*[align=c,width=0.47 \linewidth]{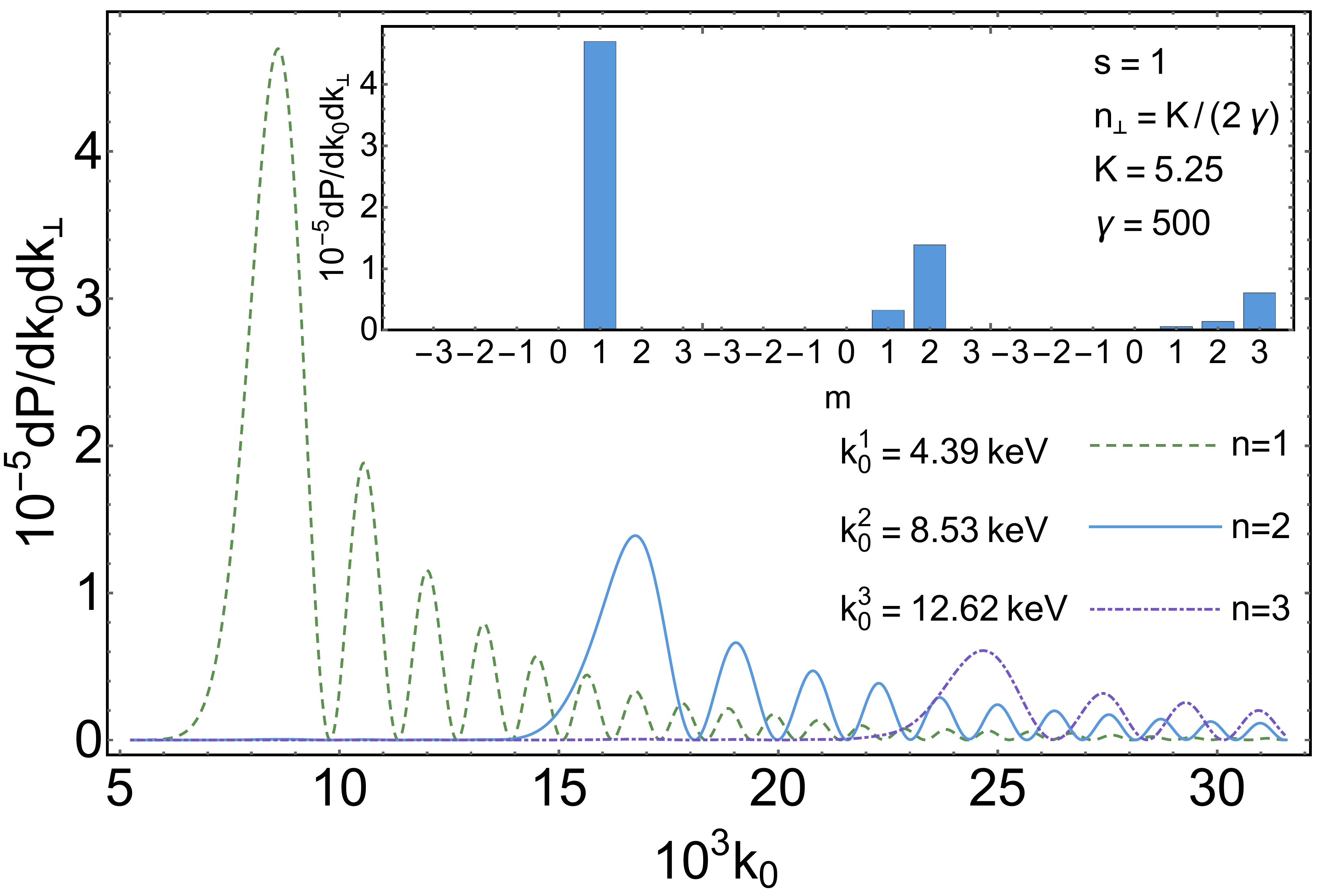}\\
\includegraphics*[align=c,width=0.47\linewidth]{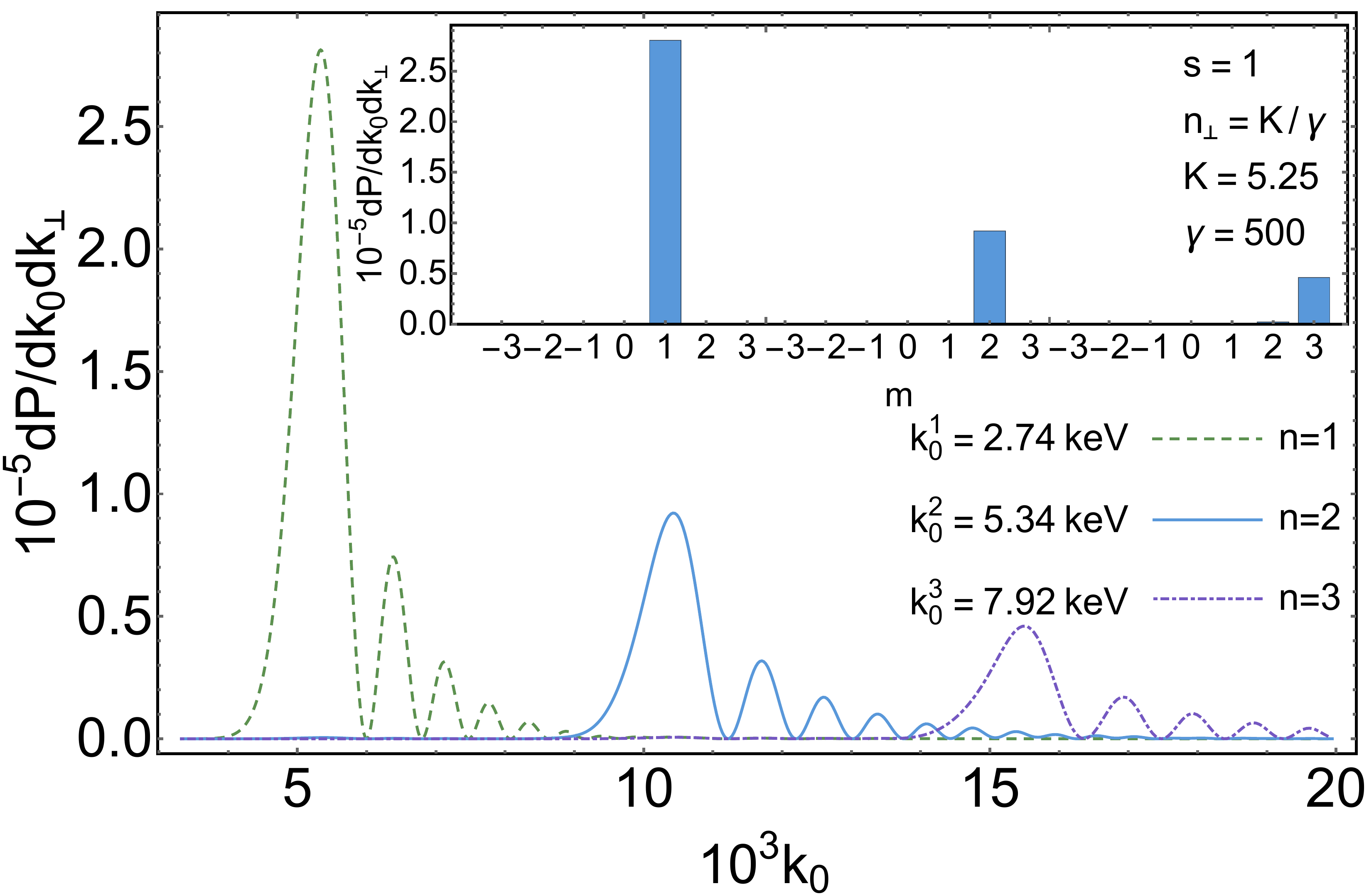}&
\includegraphics*[align=c,width=0.47\linewidth]{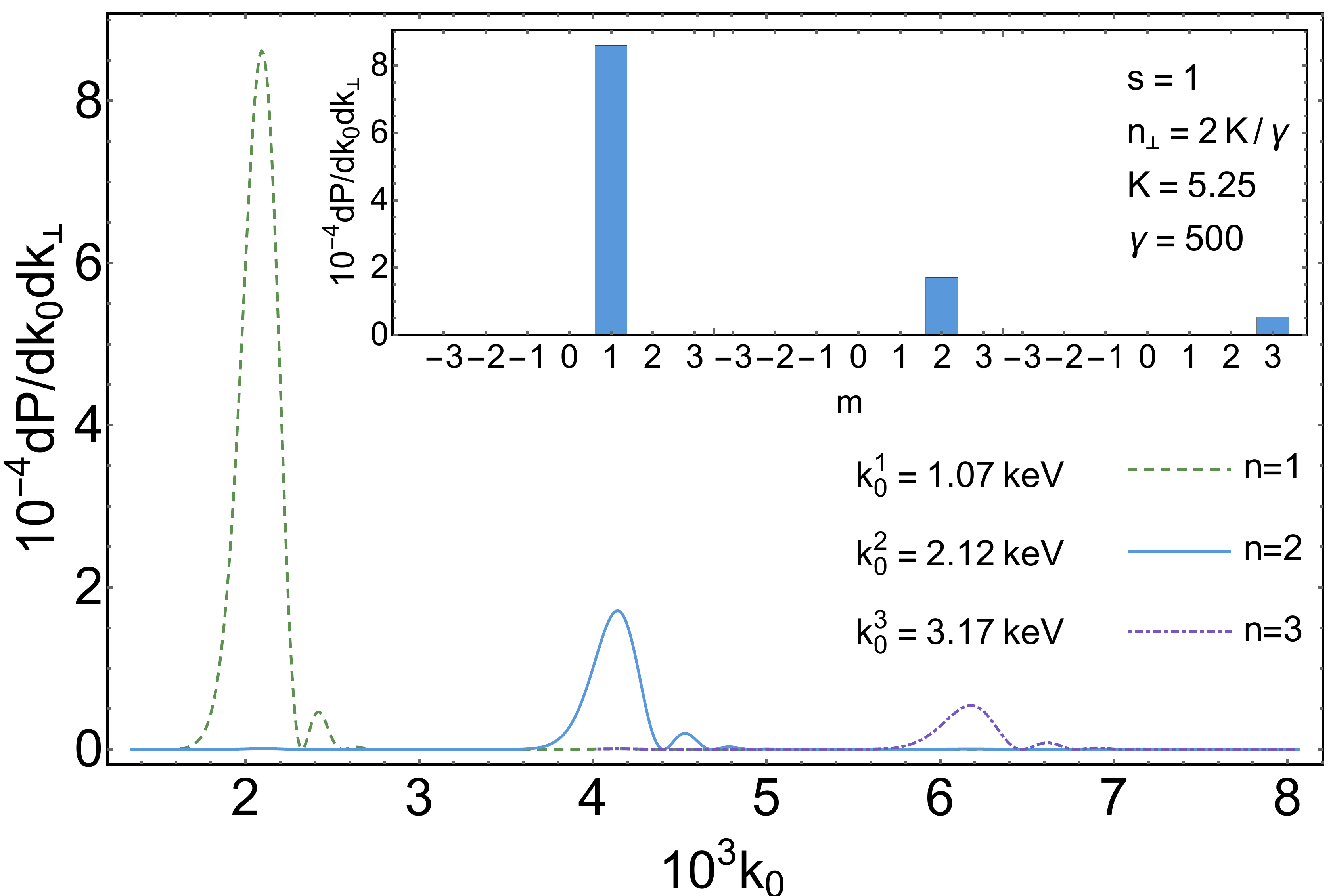}\\
\end{tabular}
    \caption{{\footnotesize The radiation of twisted photons in head-on collision of $256$ MeV electrons with the circularly polarized  electromagnetic wave produced by the CO$_2$ laser with the wavelength $10\,\mu$m, intensity $10^{18}$ W/cm${}^2$, and amplitude envelope \eqref{envelope} with $N=20$. These data correspond to $\Omega\approx2.8\times 10^{-7}$ and $a_0\approx1.47\times10^{-6}$. The first three harmonics \eqref{harmonics_new} are depicted. The applicability conditions \eqref{m_ineq} are satisfied for $\s_\perp\lesssim10^3m^{-1}$. The photon energy is measured in the electron rest energies. Inset: The distributions over $m$ at the main maxima of harmonics.}}
    \label{fig_CO2}
\end{figure}

Let us consider separately the regular, weakly degenerate, and strongly degenerate cases. In the regular case, under the same assumptions that was made in considering the first term in \eqref{prob_by_Dir}, we deduce up to an irrelevant phase
\begin{equation}
\begin{split}
    I_a\approx\frac{k_0}{4\ups'_0}\sum_{n=1}^\infty\frac{\theta(a_n)\theta(b_n)}{\Omega\sqrt{a_nb_n}}e^{in\vf_{0N}}&\Big\{ e^{in\de+im\psi+i k_\perp|z_+|\cos(\psi-\arg z_+)}g^a_n(\psi)\Big|_{\psi=\rho+\xi_n}+\\
    &+e^{in\de+im\psi +i k_\perp|z_+|\cos(\psi-\arg z_+)}g^a_n(\psi)\Big|_{\psi=\rho-\xi_n} \Big\}.
\end{split}
\end{equation}
The respective contribution to the probability, in the region of photon energies where the harmonics do not overlap, becomes
\begin{equation}
\begin{split}
    dP_a(s,m,k_\perp,k_3)\approx \frac{e^2k_0^2}{16\ups'^2_0}\sum_{n=1}^\infty \frac{\theta(a_n)\theta(b_n)}{\Omega^2a_nb_n}&\bigg|e^{in\de+im\psi +i k_\perp|z_+|\cos(\psi-\arg z_+)}g^a_n(\psi)\Big|_{\psi=\rho+\xi_n}+\\
    &+e^{in\de+im\psi +i k_\perp|z_+|\cos(\psi-\arg z_+)}g^a_n(\psi)\Big|_{\psi=\rho-\xi_n} \bigg|^2  n_\perp^3\frac{dk_3dk_\perp}{16\pi^2}.
\end{split}
\end{equation}
This expression, just as the contribution of the first term in \eqref{prob_by_Dir}, is a periodic function of $m$ with the period given in \eqref{Tm_period}.


\begin{figure}[t]
   \centering
\begin{tabular}{cc}
\includegraphics*[align=c,width=0.47\linewidth]{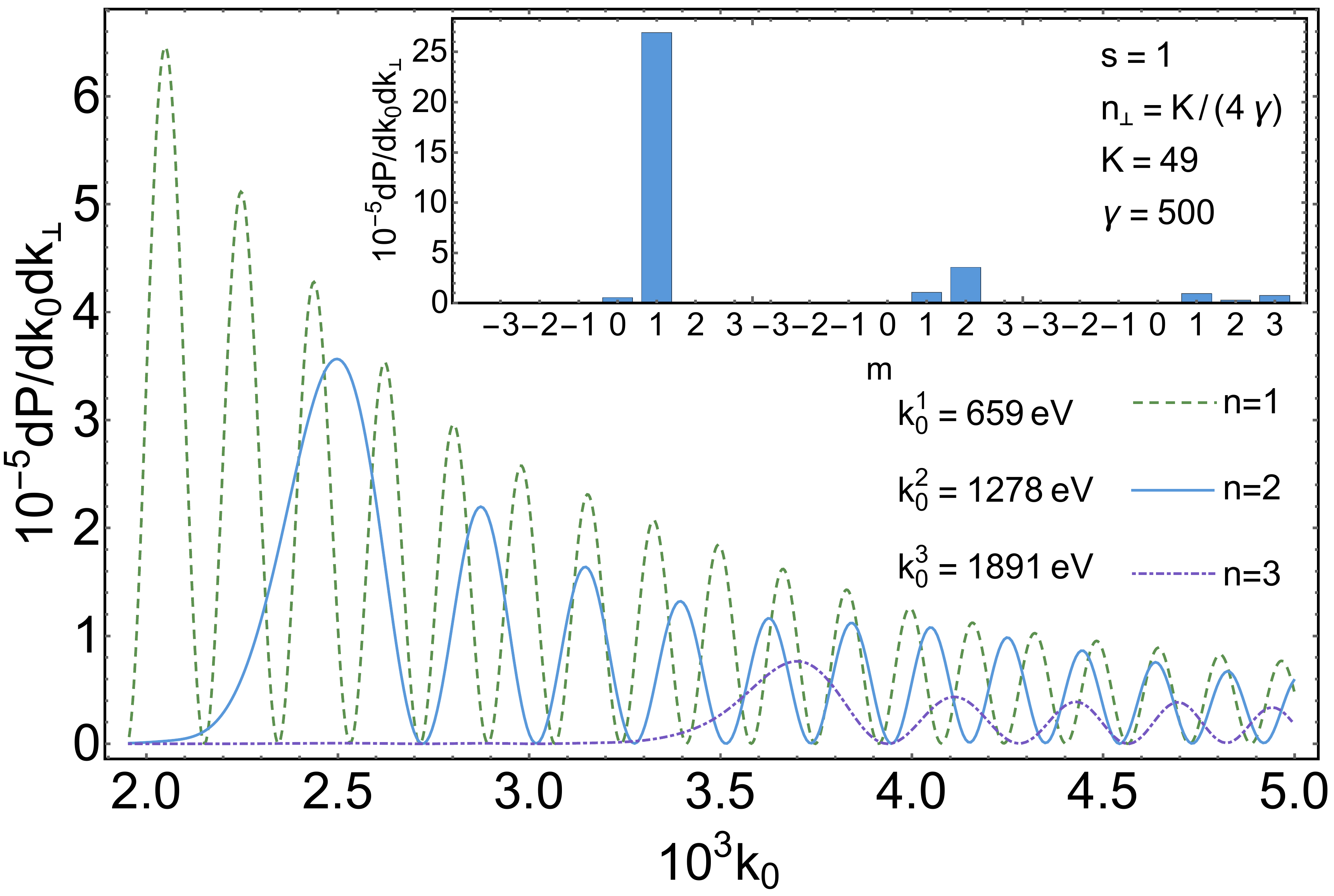}&
\includegraphics*[align=c,width=0.47 \linewidth]{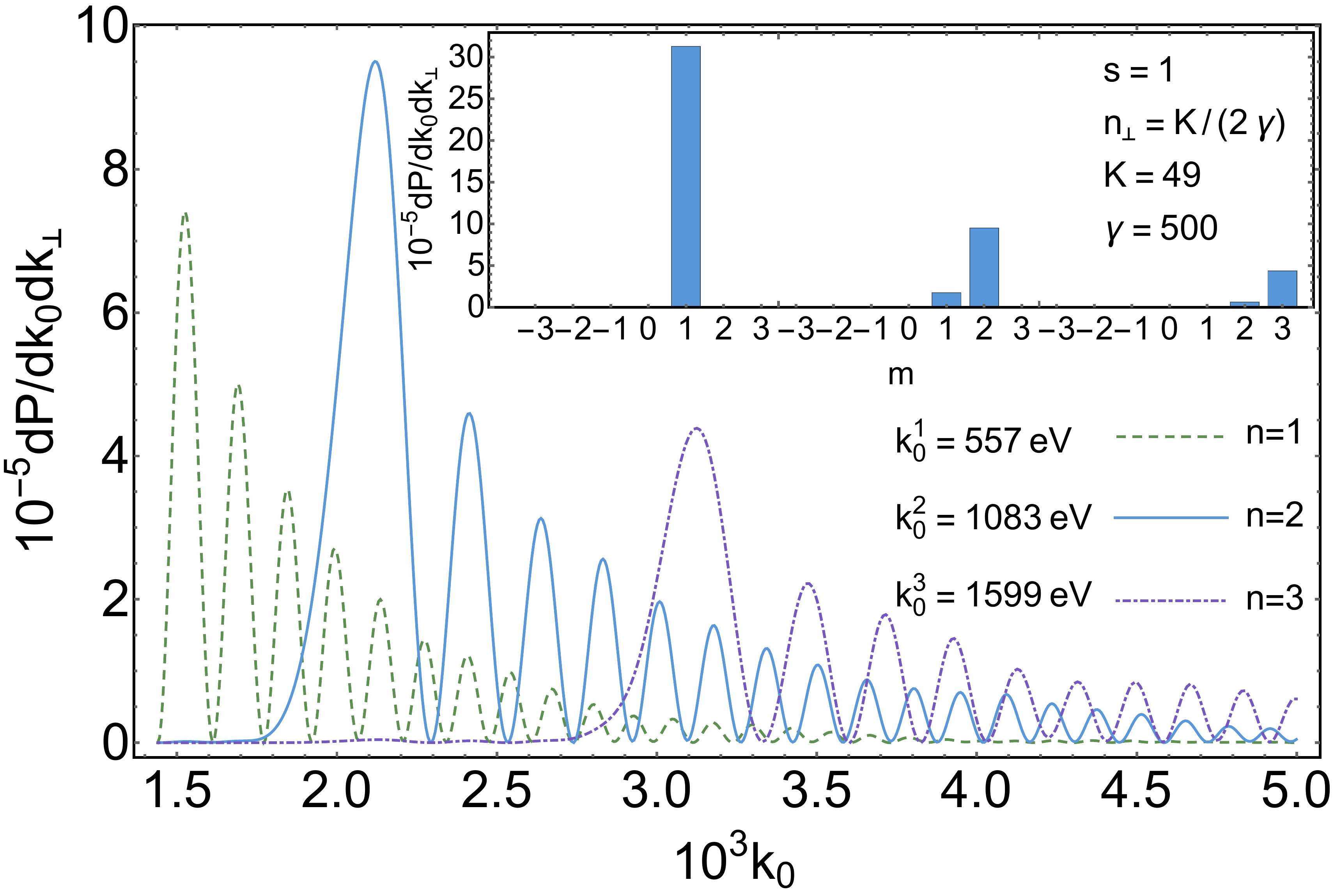}\\
\includegraphics*[align=c,width=0.47\linewidth]{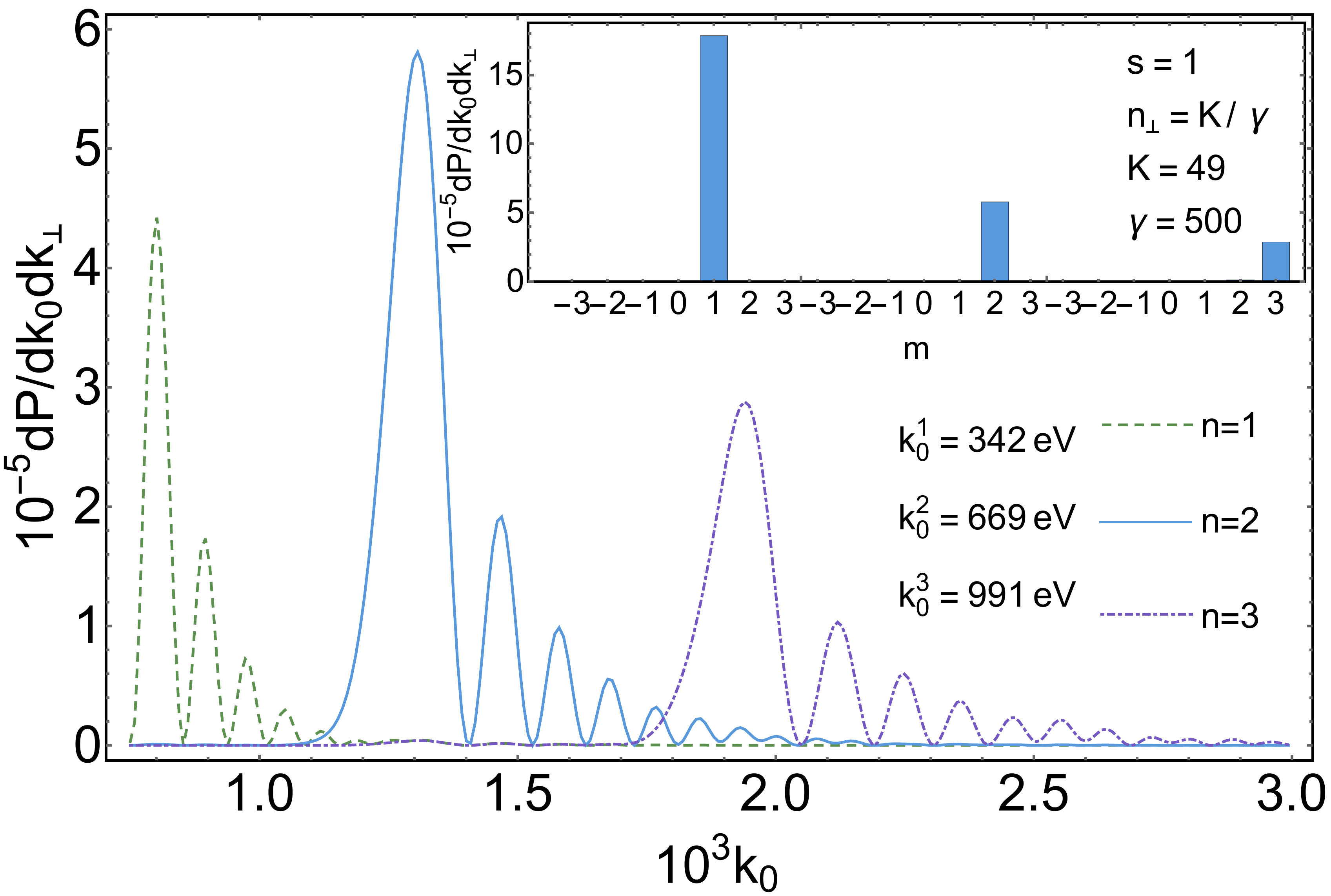}&
\includegraphics*[align=c,width=0.47\linewidth]{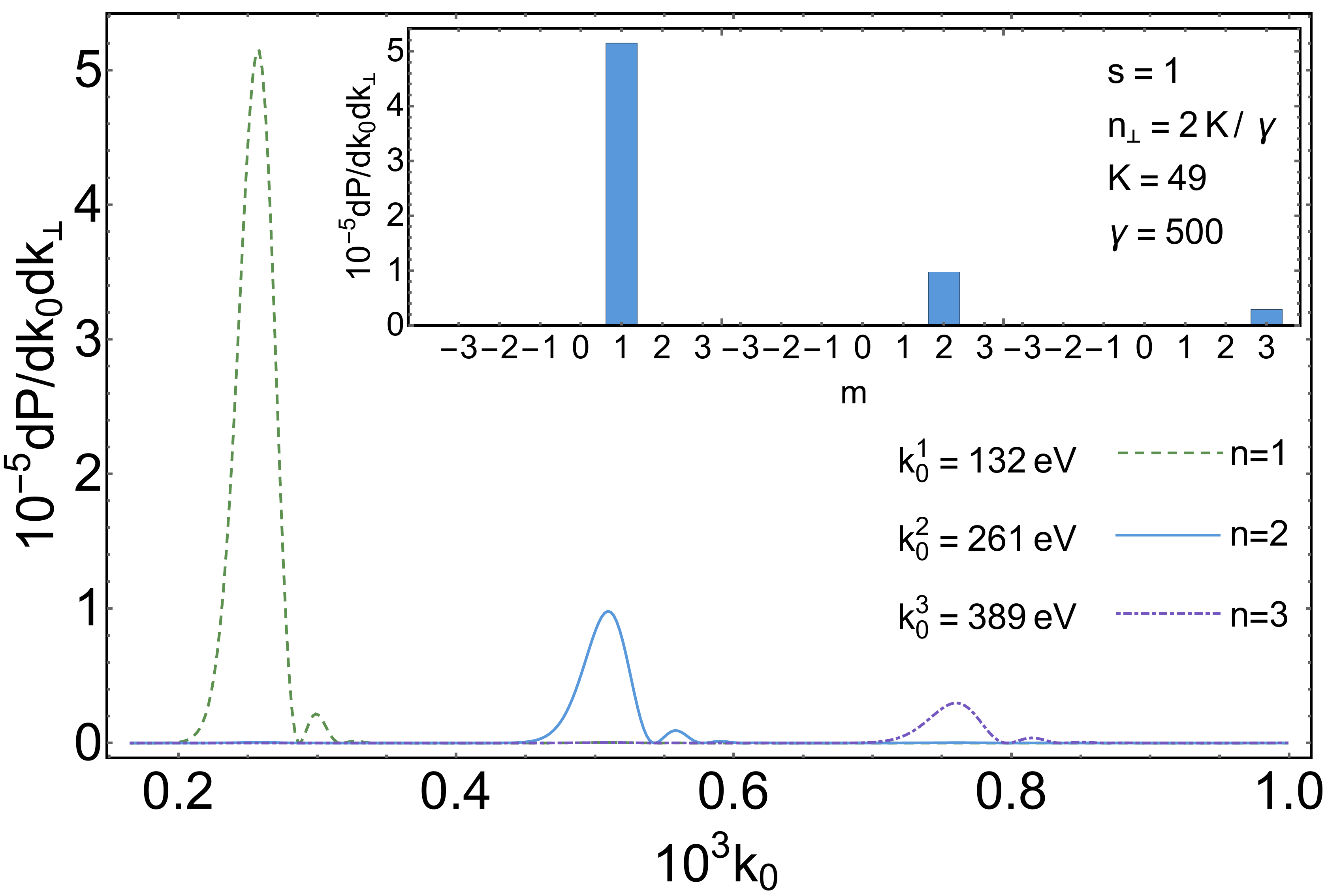}\\
\end{tabular}
    \caption{{\footnotesize The radiation of twisted photons in head-on collision of $256$ MeV electrons with the circularly polarized  electromagnetic wave produced by the Ti:Sa laser with parameters \eqref{las_wave} and amplitude envelope \eqref{envelope} with $N=20$. The first three harmonics \eqref{harmonics_new} are depicted. The applicability conditions \eqref{m_ineq} are satisfied for $\s_\perp\lesssim10^3m^{-1}$. The photon energy is measured in the electron rest energies. Inset: The distributions over $m$ at the main maxima of harmonics.}}
    \label{fig_TiSa}
\end{figure}

In the weakly degenerate case, for $a_n=0$, $b_n>0$, i.e., $\bar{k}_0=n\omega_+$, we obtain
\begin{equation}
    dP_a(s,m,k_\perp,k_3)\approx e^2 \frac{k_0^2}{\ups'^2_0} \frac{N|g_n^a(\rho)|^2n_\perp^3}{\Omega^2 n(\omega_+\omega_-^{-1}-1)} \frac{dk_3dk_\perp}{32\pi^2}.
\end{equation}
If $b_n=0$, $a_n>0$, i.e., $\bar{k}_0=n\omega_-$, then
\begin{equation}
    dP_a(s,m,k_\perp,k_3)\approx e^2 \frac{k_0^2}{\ups'^2_0} \frac{N|g_n^a(\pi-\rho)|^2n_\perp^3}{\Omega^2 n(1-\omega_-\omega_+^{-1})} \frac{dk_3dk_\perp}{32\pi^2}.
\end{equation}
These expressions are independent of $m$.

In the strongly degenerate case $a_n=0$, $b_n=0$, we suppose that $\bar{\ups}_\perp=0$. Then $x_n$ does not depend on $\psi$,
\begin{equation}
    g_n^a= J_n(\eta)-\frac{2K}{n_\perp\ups_-}J_{n-s}(\eta)= \Big(1-\frac{2n\Omega}{n_\perp k_\perp}\Big) J_n(\eta)-\frac{2sK}{n_\perp\ups_-}J'_n(\eta)=g_n,
\end{equation}
and relations \eqref{forw_chi_psi} take place. The contribution to the probability to detect a twisted photon is
\begin{equation}
    dP_a(s,m,k_\perp,k_3)=e^2\frac{k_0^2}{\ups'^2_0} \de_N^2(x_n) J_{m-n}^2(k_\perp |z_+|)g_n^2 n_\perp^3\frac{dk_3dk_\perp}{64}.
\end{equation}
The total probability \eqref{tot_prob} becomes (cf. \eqref{prob_by_Dir_wigg})
\begin{equation}\label{forw_laser_rec}
    dP(s,m,k_\perp,k_3)=e^2(1+q^2) \de_N^2(x_n) J_{m-n}^2(k_\perp |z_+|)g_n^2 n_\perp^3\frac{dk_3dk_\perp}{32}.
\end{equation}
For $k_\perp|z_+|\ll1$, the selection rule $m=n$ is fulfilled. As in the case of forward undulator radiation, the applicability conditions \eqref{1photon}, \eqref{m_ineq} must be satisfied. The number of twisted photons recorded by the detector is approximately given by \eqref{phot_raded}.


\begin{figure}[t]
   \centering
\includegraphics*[align=c,width=0.48\linewidth]{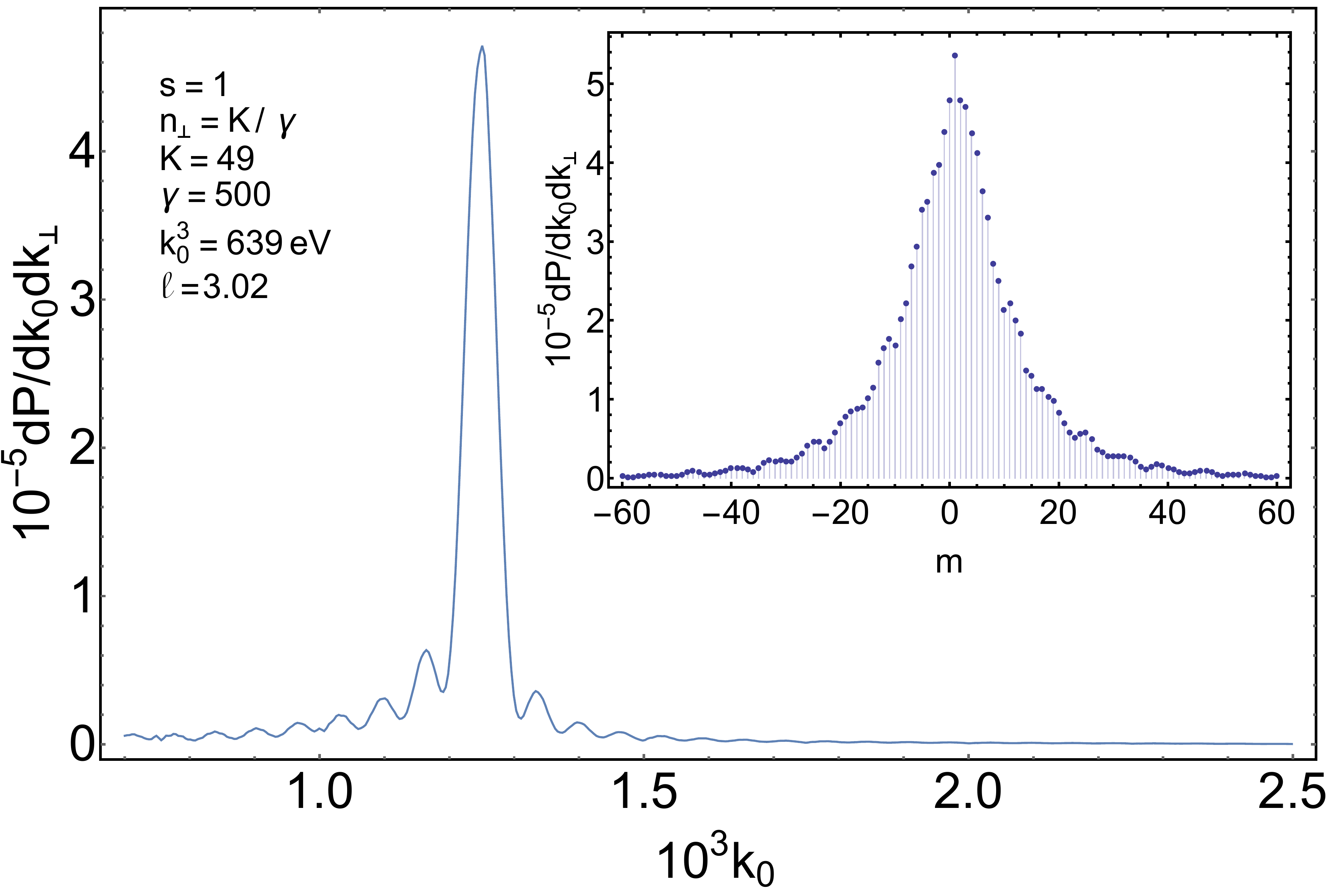}\;
\includegraphics*[align=c,width=0.49\linewidth]{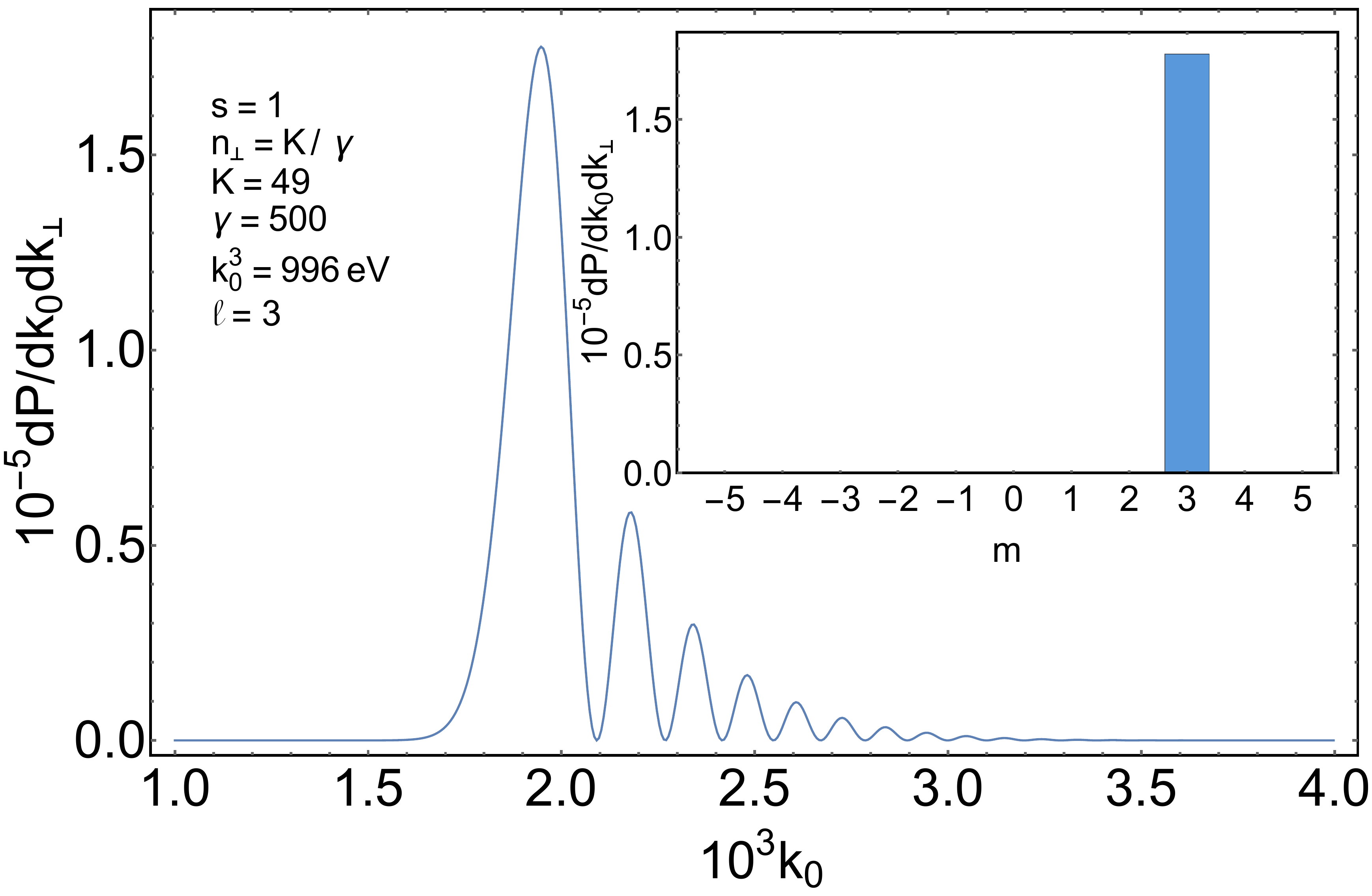}
    \caption{{\footnotesize The radiation of twisted photons in head-on collision of $256$ MeV electrons with the circularly polarized  electromagnetic wave produced by the Ti:Sa laser. The parameters are the same as in Fig. \ref{fig_TiSa} but the laser pulse amplitude envelopes are different. Left panel: The amplitude envelope has the form \eqref{envelope_const}. The initial laser wave phase $\vf_0=0$. The maximum near $k_0=3\omega_+$ is shown. Right panel: The amplitude envelope is $a_0\sin^4(\Omega\xi/(2N))$. The third harmonic \eqref{harmonics_new} is depicted. Insets: The distributions over $m$ in the main maxima.}}
    \label{fig_Evelop}
\end{figure}

\paragraph{Examples.}

\begin{figure}[t]
   \centering
   \includegraphics*[align=c,width=0.98\linewidth]{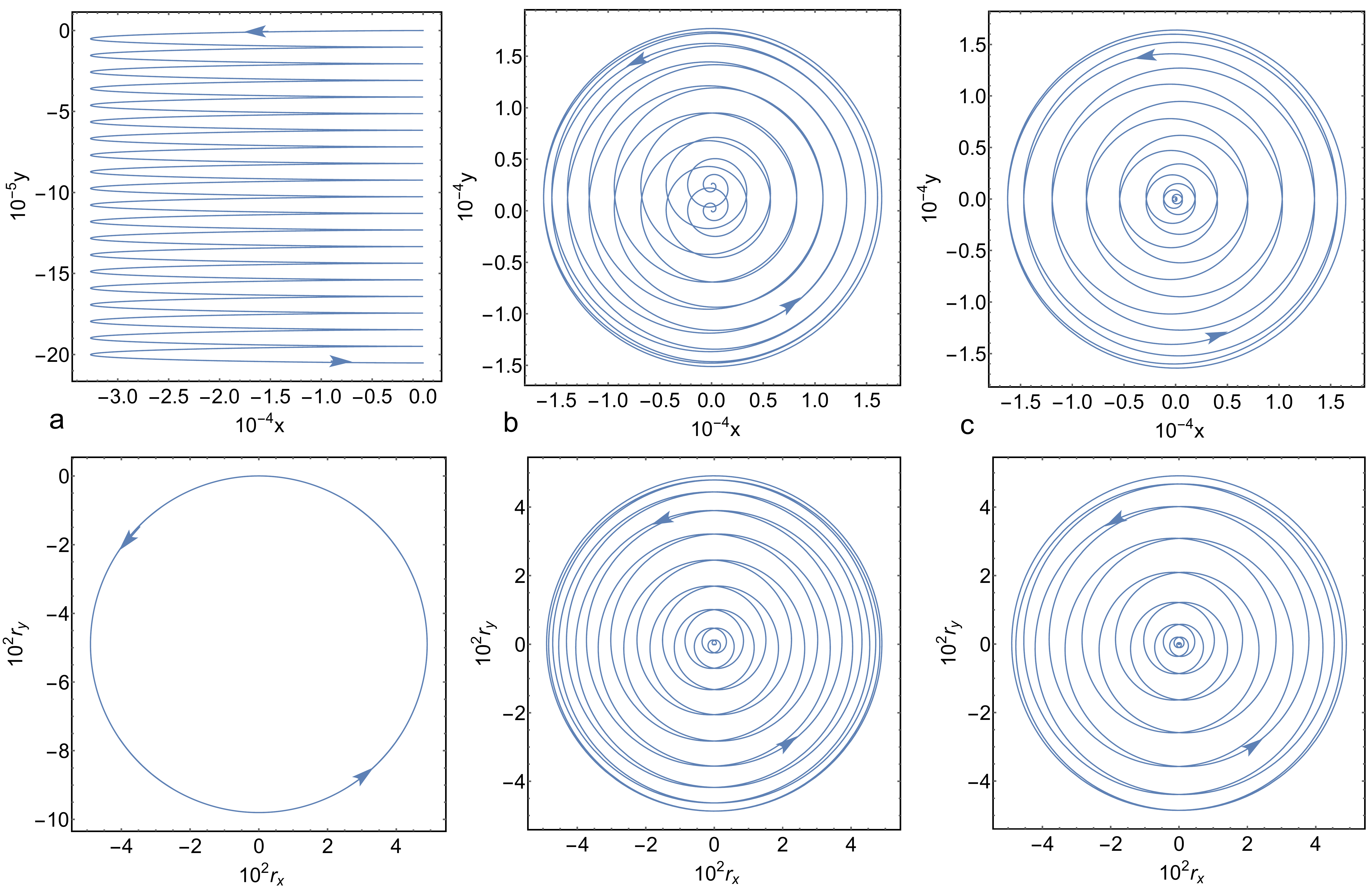}
   \caption{{\footnotesize The coordinates $x$, $y$ and ``velocities'' $r_x$, $r_y$ for head-on collision of an electron with the circularly polarized electromagnetic wave produced by the Ti:Sa laser. The parameters are the same as in Fig. \ref{fig_Evelop}. The electron moves initially along the detector axis. The initial laser wave phase $\vf_0=0$. The lengths are measured in the Compton wavelengths. a) The amplitude envelope has the form \eqref{envelope_const}. b) The amplitude envelope is given in \eqref{envelope}. c) The amplitude envelope is $a_0\sin^4(\Omega\xi/(2N))$.}}
    \label{fig_Traj}
\end{figure}

As is seen from \eqref{forw_laser}, \eqref{forw_laser_rec}, the electrons moving in the laser wave represent a pure source of twisted photons only in the strongly degenerate case when $\bar{\ups}_\perp\approx0$ and $k_\perp|z_+|\approx0$. For this to be the case, the initial data must be taken in the form (see \eqref{r_plane_wave_1}, \eqref{x_bar}, \eqref{zpm})
\begin{equation}\label{strong_deg}
    r_+(0)=iK\ups_-^{-1}e^{i\vf_0},\qquad x_+(0)=K\ups_-^{-1}\Omega^{-1}e^{i\vf_0}.
\end{equation}
Since it is very hard to control the initial phase of a laser wave, in the wiggler regime, $K\gtrsim3$, it is practically impossible to launch the electron to the electromagnetic wave so that the radiation produced by it would correspond to the strongly degenerate case. For an arbitrarily chosen phase, equalities \eqref{strong_deg} are strongly violated, harmonics \eqref{spectrum_laser} spread violently, and the twisted photon detector records a wide distribution over $m$ (see Fig. \ref{fig_Evelop}). In this case, the twisted photons escape the laser wave at large angles to the detector axis rather than move along it (see Fig. \ref{fig_Traj}).

It turns out that this situation can be improved if one takes into account that, usually, the laser wave pulses generated in experiments have no sharp rising and descending edges. The amplitude envelope $a(\xi)$ is a smooth function vanishing at $\xi=\{0,TN\}$ and $a'(\xi)/a(\xi)\sim1/N$. For $N\gtrsim10$, this entails that the radiation probability ceases to depend virtually on the initial phase $\vf_0$. As a result, it is possible to choose the initial velocity and entrance point of the electron to the electromagnetic wave so that the corresponding radiation will be a sufficiently pure source of twisted photons.

Indeed, under the restrictions on the form of the envelope mentioned above, the integrals entering into general solution \eqref{Lorentz_sol}, \eqref{Lorentz_sol_1} of the Lorentz equations, can be approximately evaluated for $N\gtrsim10$. Integrating once by parts and keeping only the integrated term, we obtain (cf. \eqref{r_plane_wave}, \eqref{x_sol})
\begin{equation}\label{x_sol1}
\begin{gathered}
    x_\pm\approx x_\pm(0) +r_\pm(0)\xi+\frac{K}{\ups_-\Omega}e^{\pm i\vf},\qquad
    x^0\approx\frac1{2\ups_-^2}\big[(\ups_-^2+1+K^2+\ups_\perp^2(0))\xi +2\ups_\perp(0)\frac{K}{\Omega}\cos(\vf-\rho) \big],\\
    x^3\approx\frac{\zeta}{2\ups_-^2}\big[(1+K^2+\ups_\perp^2(0)-\ups_-^2)\xi +2\ups_\perp(0)\frac{K}{\Omega}\cos(\vf-\rho) \big],
\end{gathered}
\end{equation}
where $K:=a(\xi)/\Omega$ and
\begin{equation}\label{r_plane_wave1}
\begin{gathered}
    r_\pm\approx r_\pm(0) \pm iK\ups_-^{-1}e^{\pm i\vf},\qquad r^0\approx\frac1{2\ups_-^2}\big[\ups_-^2+1+K^2+\ups_\perp^2(0) -2K \ups_\perp(0)\sin(\vf-\rho) \big],\\
    r^3\approx\frac{\zeta}{2\ups_-^2}\big[1+K^2 +\ups_\perp^2(0) -\ups_-^2 -2K\ups_\perp(0)\sin(\vf-\rho) \big].
\end{gathered}
\end{equation}
The accuracy of this approximation increases as $N$ increases. As is seen, the form of the trajectory is almost the same as in the case of a laser wave with constant amplitude but without strong dependence on the initial phase. Now the dependence on the initial phase is contained only in $\vf$.


Substituting approximate trajectory \eqref{x_sol1}, \eqref{r_plane_wave1} into \eqref{I_3pm}, it is easy to see that the pure source of twisted photons can be obtained when
\begin{equation}
    x_+(0)\approx0,\qquad\ups_\perp(0)\approx0,
\end{equation}
i.e., in the strongly degenerate case. In this paper we will investigate only this case. The plots of typical trajectories in this case are presented in Fig. \ref{fig_Traj}. The calculations are performed along the same lines as those made above, except that the shift $\xi\rightarrow\xi+TN/2$ is unnecessary. In particular, formulas \eqref{forw_chi_psi}, \eqref{xn_gn_forw_las} hold and the radiation amplitude is proportional to
\begin{equation}
    I_3+\frac{1}{2}(I_++I_-)=\frac{1}{2}\sum_{n=-\infty}^\infty j_{m-n}(k_\perp x_+(0),k_\perp x_-(0))e^{in\vf_0}\int_0^{TN}d\xi g_n(\xi)e^{-i\xi x_n(\xi)}.
\end{equation}
On stretching the variable $\xi\rightarrow TN\xi$, it is clear that the integral over $\xi$ can be approximately evaluated by the WKB method. For the envelope with one maximum as, for example,
\begin{equation}\label{envelope}
    a(\xi)=a_0\sin^2(\Omega\xi/(2N)),
\end{equation}
the function $\xi x_n(\xi)$ has two extrema, $\xi_\pm(k_0)$, on the interval $(0,TN)$, in a general position. At these extrema,
\begin{equation}
    k_0^n=\frac{2\Omega n\ups_-^2}{(1-\zeta n_3)(1+K^2+2\xi KK')+(1+\zeta n_3)\ups_-^2}>0.
\end{equation}
Therefore, we have
\begin{equation}\label{amplitude_env}
    I_3+\frac{1}{2}(I_++I_-)\approx \sqrt{\frac{\pi}{2}}\sum_{n=-\infty}^\infty j_{m-n}(k_\perp x_+(0),k_\perp x_-(0))e^{in\vf_0}\Big[\frac{g_n(\xi)e^{-i\xi x_n(\xi)}}{\sqrt{i(\xi x_n(\xi))''}}\Big|_{\xi=\xi_+} +\frac{g_n(\xi)e^{-i\xi x_n(\xi)}}{\sqrt{i(\xi x_n(\xi))''}}\Big|_{\xi=\xi_-} \Big],
\end{equation}
where the principal branch of the square root is taken. The contribution of the boundaries is suppressed since, in the leading order in $1/N$, the same contribution but with opposite sign comes from the edge radiation (see for details, e.g., \cite{BKL1,BKL3}). As a result, the contributions of the internal stationary points are only relevant. One of the extremum points, $\xi_+$, is close to the point where $a'(\xi)=0$, i.e., $K(\xi)$ is close to its maximum value at this point. Taking into account the form of $g_n(\xi)$, we see that this stationary point gives the leading contribution to \eqref{amplitude_env}. The main maximum is located approximately at
\begin{equation}\label{harmonics_new}
    k_0^n\approx\frac{2\Omega n\ups_-^2}{(1-\zeta n_3)(1+K^2_{max})+(1+\zeta n_3)\ups_-^2},\qquad\Omega n>0.
\end{equation}
In fact, the maximum is slightly shifted to the right since the stationary point $\xi_+$ is displaced a little from the extremum of $K(\xi)$. If $k_\perp|x_+(0)|\ll1$, then $j_{m-n}(k_\perp x_+(0),k_\perp x_-(0))=\de_{mn}$ and
\begin{equation}\label{prob_laser}
    dP(s,m,k_\perp,k_3)=e^2\bigg|\frac{g_m(\xi)e^{-i\xi x_m(\xi)}}{\sqrt{i(\xi x_m(\xi))''}}\Big|_{\xi=\xi_+} +\frac{g_m(\xi)e^{-i\xi x_m(\xi)}}{\sqrt{i(\xi x_m(\xi))''}}\Big|_{\xi=\xi_-} \bigg|^2n_\perp^3\frac{dk_3dk_\perp}{32\pi}.
\end{equation}
At the extremum points $\xi\sim N$, $x'_n(\xi)\sim 1/N$, and $x''_n(\xi)\sim 1/N^2$. Therefore, the radiation probability is proportional to $N$.

\begin{figure}[t]
   \centering
\includegraphics*[align=c,width=0.48\linewidth]{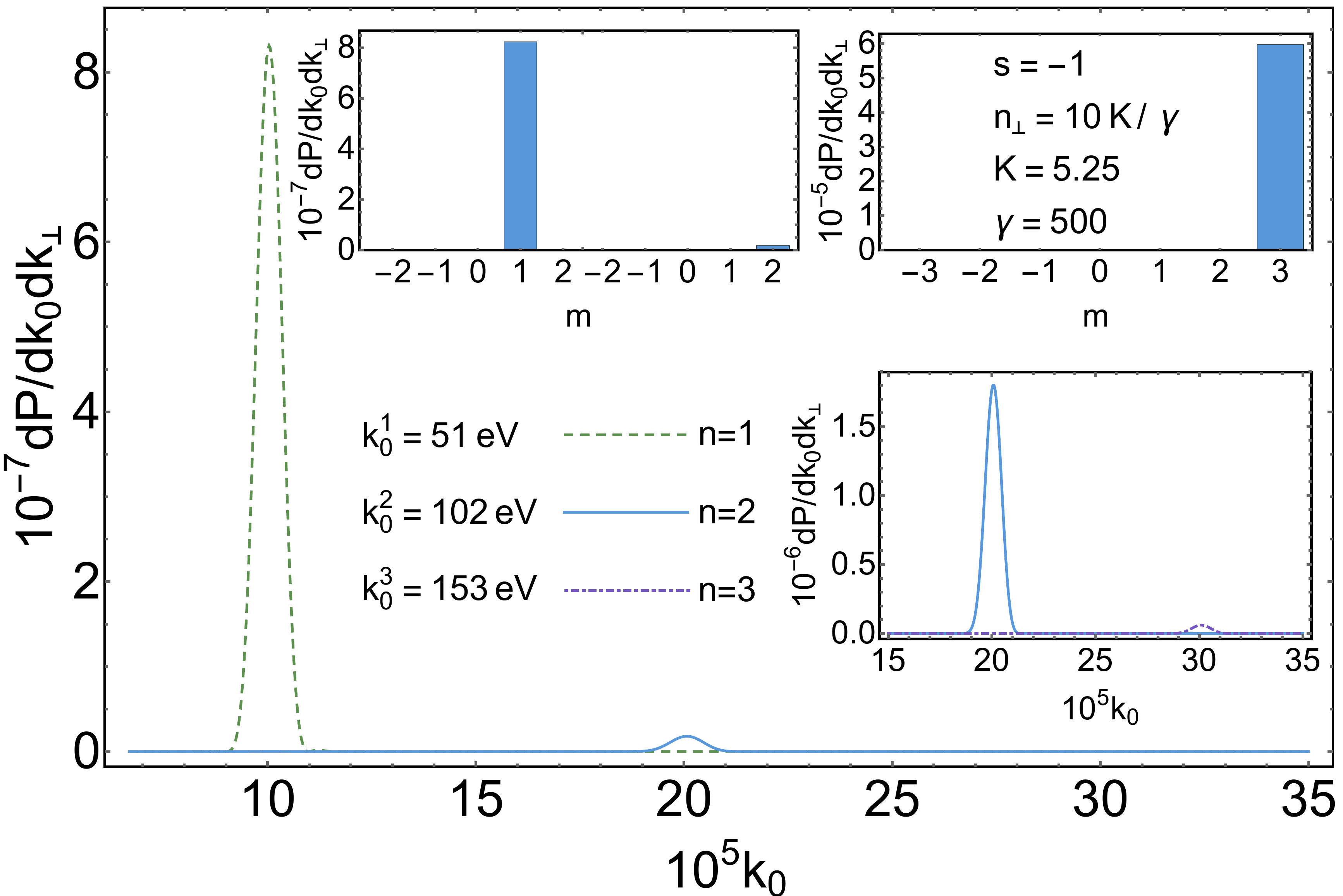}\;
\includegraphics*[align=c,width=0.48\linewidth]{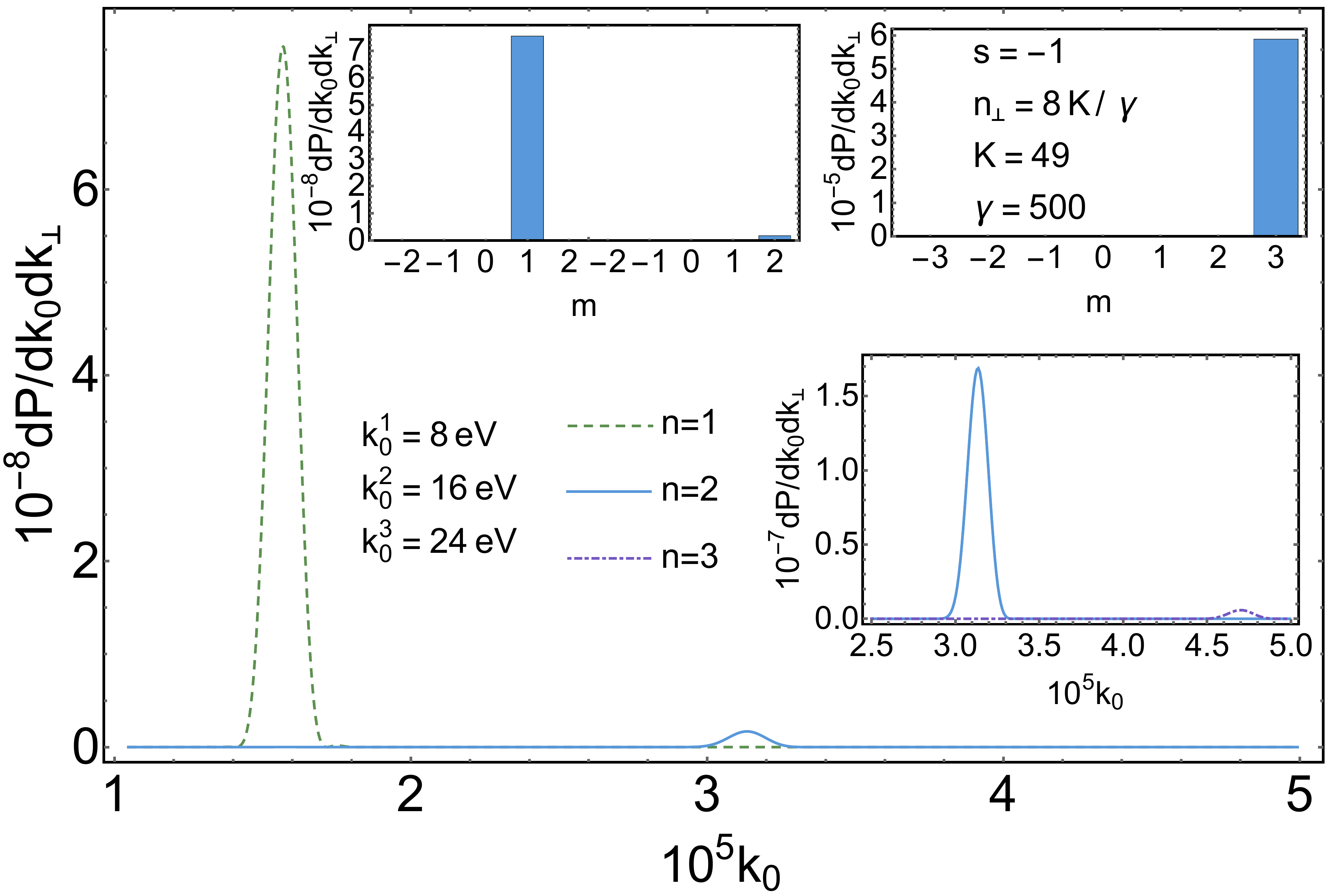}
    \caption{{\footnotesize The imprint of blossoming out rose effect \cite{Bord.1} on the radiation of twisted photons in head-on collision of electrons with circularly polarized electromagnetic wave produced by the CO$_2$ (left panel) and Ti:Sa (right panel) lasers. The parameters are the same as in Figs. \ref{fig_CO2}, \ref{fig_TiSa}. The first harmonic \eqref{harmonics_new} with $s=-1$ dominates. This is just a Lorentz boosted back lobe of the first harmonic of synchrotron radiation (see Sec. 1.3.4 of \cite{Bord.1}). It is mostly left-handed polarized, and this property is preserved by the Lorentz transformations. This harmonic does not die out even for $n_\perp\approx1$. Upper left inset: The distributions over $m$ at the main maxima of the first two harmonics. Upper right inset: The distribution over $m$ at the main maximum of the third harmonic. Lower inset: The second and third harmonics are separately depicted.}}
    \label{fig_Rose}
\end{figure}

The dependence of the probability density on $k_0$ differs from the profile $\de_N^2(x_m)$ and is depicted in Figs. \ref{fig_CO2}, \ref{fig_TiSa}, \ref{fig_Rose}. As is seen, the radiation probability $dP(m)$ is nearly zero for $k_0<k_0^m$, then it rapidly grows in the vicinity of $k_0=k_0^m$, and for $k_0>k_0^m$ it declines performing oscillations to zero. When $n_\perp\ga \lesssim K_{max}/2$, this decrease is quite slow. So, for these values of $n_\perp$, the radiation probability $dP(m,k_0)$ taken in the neighborhood of the point $k_0=k_0^m$ can contain a considerable contribution of photons with the projection of the total angular momentum $m-1$. For very small $n_\perp$, the contribution of photons with all the lower projections of the angular momentum are relevant (see Figs. \ref{fig_CO2}, \ref{fig_TiSa}). When $n_\perp\ga\approx K_{max}$, the peaks of $dP(m,k_0)$ with different $m$ are virtually not overlapping. In that case, the radiation at $k_0\approx k_0^m$ consists of twisted photons with the projection of the angular momentum $m$, i.e., the selection rule $m=\chi n$ is fulfilled, where $n$ is the number of harmonic \eqref{harmonics_new} and $\chi=\pm1$ is the handedness of the helix along which the electron is moving.

As follows from \eqref{phot_raded}, the most part of twisted photons is radiated at lower harmonics \eqref{harmonics_new}. In the wiggler case, $K\gg1$, these harmonics are fairly well described as the Lorentz boosted lower harmonics of synchrotron radiation (see the discussion after \eqref{dPko}). They would perfectly coincide if the charged particle moved along an ideal helix. Those lower harmonics were studied in \cite{Bord.1} where the effect of blossoming out rose was established: in the ultrarelativistic limit, even for $\be=1$, these harmonics do not drive to the orbit plane, and the maximum intensity of radiation of every harmonic is achieved at some finite angle to the orbit plane. These angles for $\be=1$ are given in Sec. 1.3.4 of \cite{Bord.1}. The maxima of the first harmonic are located at the angles $\theta'=\{0,\pi\}$.

In the laboratory frame, the imprint of this effect on the properties of radiation is as follows. One observes the maxima of radiation of twisted photons at these harmonics at the angles taken from \cite{Bord.1} and substituted into \eqref{Lorentz_ang}. The orbit plane is seen in the laboratory frame as a cone with opening $2K/\ga$. Since, in the synchrotron frame, the front lobes are mostly right-handed polarized (if, in the laboratory frame, the particle moves along a right-handed helix) and the back lobes are mostly left-handed polarized, and this property is Lorentz invariant, in the laboratory frame, the twisted photons with $s=1$ dominate for $n_\perp<K/\ga$ while, for $n_\perp>K/\ga$, the twisted photons with $s=-1$ prevail. The first harmonic with $s=-1$ does not die out for large $n_\perp$ and even for $n_\perp\approx1$. The plots of lower harmonics are presented in Fig. \ref{fig_Rose}. Strictly speaking, formula \eqref{prob_by_Dir} is not applicable for so large $n_\perp$. However, in the case of small quantum recoil, we can use exact formula [(36), \cite{BKL2}]. Numerical calculation shows that formulas \eqref{prob_by_Dir} and [(36), \cite{BKL2}] give the same results in this case.


\section{Conclusion}

Let us summarize the results. Using the BK semiclassical approach \cite{BaiKat1,BKMS,BaKaStr,BaKaStrbook}, we derived the general formula for the one-photon radiation probability of a twisted photon by scalar \eqref{prob_by_scal} and Dirac \eqref{prob_by_Dir} ultrarelativistic particles moving in the electromagnetic field of a general configuration. This formula takes into account the quantum recoil undergone by a charged particle in radiating the twisted photon and, in the case of negligible recoil, turns into the formula given in \cite{BKL2}. Then we applied this formula to radiation of charged particles in helical undulators and in circularly polarized laser waves with a plane wavefront. The explicit formulas for the probability to record the twisted photon by a detector were obtained in these cases. The inclusion of quantum recoil forbids radiation of twisted photons with energies larger than the initial particle energy. We established that, as a rule, the quantum recoil increases the total yield of radiation in comparison with classical formulas (see Fig. \ref{fig_FEL}) and, at the same time, diminishes the energy of radiated photons. The spin degrees of freedom of a radiating particle increase the probability of radiation of twisted photons.

The conditions when the developed semiclassical approach is justified were found and analyzed. The most stringent among these conditions is \eqref{trans_coh}. It guarantees that, in describing the radiation of twisted photons, it is sufficient to characterize the particle wave packet by its average coordinate and momentum. In particular, it turns out that the radiation of twisted photons with large projection $m$ of the total angular momentum produced by electrons in helical wigglers and strong laser waves can be described semiclassically only in the case of a small quantum recoil. In the dipole regime, the quantum recoil can be substantial and still be described semiclassically (see Fig. \ref{fig_FEL}). We found estimate \eqref{phot_raded} for the number of twisted photons with large projections of the total angular momentum produced in the forward radiation. We also described the effect of blossoming out rose \cite{Bord.1} in the radiation of twisted photons by electrons evolving in strong laser waves with circular polarization and wigglers (see Fig. \ref{fig_Rose}).

As an example, we considered the radiation of twisted photons with large angular momentum in the helical wiggler (see Fig. \ref{fig_WIGG}) and in the circularly polarized strong laser waves produced by the CO${}_2$ and Ti:Sa lasers (Figs. \ref{fig_CO2}, \ref{fig_TiSa}, \ref{fig_Evelop}, \ref{fig_Rose}). The parameters are given in these figures. In particular, we showed that MeV twisted photons with $m\sim5$ can be generated in helical wigglers. As for lasers, we found that the design of a sufficiently pure source of twisted photons based on the nonlinear Compton process is only possible for long laser pulses, $N\gtrsim10$, with a smooth amplitude envelope. For short pulses, the escape direction of a twisted photon depends severely on the initial phase of a laser wave that is virtually uncontrollable. Therefore, the detector (the atom, for example) will feel the radiation consisting of twisted photons with wide spread of the total angular momentum projections $m$ (see Fig. \ref{fig_Evelop}). For the systems concerned, we also described the effect of a finite width of a particle bunch on the incoherent radiation of twisted photons \cite{BKb}.

\paragraph{Acknowledgments.}

We are thankful to Yu.L. Pivovarov and D.V. Karlovets for fruitful conversations. This work is supported by the Russian Science Foundation (project No. 17-72-20013).

\appendix
\section{Twisted photons in terms of plane waves}\label{TwPlw}

For the convenience of the reader, we shall provide the representation of the states of twisted photons in terms of the plane wave ones (see the detailed exposition, e.g., in \cite{JenSerepj}). The states describing the photons with plane wave front,
\begin{equation}\label{pl_wav_st}
    |s,k_1,k_2,k_3\ran,
\end{equation}
and the states of twisted photons,
\begin{equation}\label{tw_phot_st}
    |s,m,k_\perp,k_3\ran,
\end{equation}
constitute the complete sets in the Hilbert space of one-particle states:
\begin{equation}
\begin{split}
    &\sum_{s=\pm1}\int\frac{Vd\spk}{(2\pi)^3}2k_0 |s,k_1,k_2,k_3\ran\lan s,k_1,k_2,k_3|=\\
    &= \sum_{s=\pm1}\sum_{m=-\infty}^\infty \int_{-\infty}^\infty\frac{L_3dk_3}{2\pi}\int_0^\infty\frac{Rdk_\perp}{\pi} 2k_0 |s,m,k_\perp,k_3\ran\lan s,m,k_\perp,k_3|=1.
\end{split}
\end{equation}
The states are normalized as
\begin{equation}
    \lan s,k_1,k_2,k_3|s,k_1,k_2,k_3\ran=\lan s,m,k_\perp,k_3| s,m,k_\perp,k_3\ran=(2k_0)^{-1}
\end{equation}
where $k_0=|\spk|=\sqrt{k_\perp^2+k_3^3}$. One can decompose the state \eqref{tw_phot_st} in terms of the states \eqref{pl_wav_st}. Carrying out rather simple calculations, we come to
\begin{equation}\label{tw_w_in_pl_w}
    |s,m,k_\perp,k_3\ran=\frac{\sqrt{2k_0V}}{2\sqrt{R L_3}}\Big(\frac{k_\perp}{k_0}\Big)^{1/2}\int_{-\pi}^\pi\frac{d\vf}{2\pi} i^{-m}e^{im\vf}|s,k_\perp\cos\vf,k_\perp\sin\vf,k_3\ran.
\end{equation}

\section{Evaluation of integrals over the azimuth angle}\label{Azim_Int}

It is convenient to evaluate the integrals over the azimuth angles of the vectors $\spk_{1,2}$ in expression \eqref{AA_BB} as follows. Up to a common factor, which can be restored easily from \eqref{tw_w_in_pl_w}, we have the correspondence
\begin{equation}
    \mathbf{f}^*_1\rightarrow(\tfrac12[a^*_+\mathbf{e}_+ +a^*_-\mathbf{e}_-]+a^*_3\mathbf{e}_3)e^{-ik_3 q_1 x_{13}}=:\mathbf{a}^*_1,\qquad \mathbf{f}_2\rightarrow(\tfrac12[a_+\mathbf{e}_- +a_-\mathbf{e}_+]+a_3\mathbf{e}_3)e^{ik_3q_2x_{23}}=:\mathbf{a}_2.
\end{equation}
Hereinafter, for brevity, we write only those arguments of the mode functions $a_{\pm,3}$, $a^*_{\pm,3}$ that differ from those written in formula \eqref{a_pm3}. The basis vectors $\spe_i$ are defined in [(9), \cite{BKL2}]. Notice that $q_{1,2}=q_i$ in \eqref{AA_BB} but we keep $q_{1,2}$ different. Then, for example,
\begin{equation}
    (\mathbf{f}^*_1\dot{\mathbf{x}}_1)(\mathbf{f}_2\dot{\mathbf{x}}_2)\rightarrow e^{ik_3(q_2 x_{23} - q_1x_{13})}(\tfrac12[\dot{x}_{1-} a^*_- +\dot{x}_{1+} a^*_+]+\dot{x}_{13}a^*_3) (\tfrac12[\dot{x}_{2+} a_- +\dot{x}_{2-} a_+]+\dot{x}_{23}a_3)=(\mathbf{a}^*_1\dot{\spx}_1)(\mathbf{a}_2\dot{\spx}_2).
\end{equation}
The additional powers of $\spk_{1,2}$ can be obtained by differentiation of the expression with respect to
\begin{equation}
    \mathbf{b}_{1,2}:=q_{1,2}x_{3\,1,2}\spe_3+\spx_{\perp1,2}.
\end{equation}
For example,
\begin{equation}
    k_0(\mathbf{f}_2\mathbf{f}^*_1)(\dot{\mathbf{x}}_2\spk_1)\rightarrow ik_0\Big(\dot{x}_{23}\frac{\partial}{\partial b_{13}} +\dot{x}_{2+}\frac{\partial}{\partial b_{1+}} +\dot{x}_{2-}\frac{\partial}{\partial b_{1-}}\Big)  (\mathbf{a}_1^*\mathbf{a}_2).
\end{equation}
The derivatives of the mode functions are calculated with the aid of relations [(A3), \cite{BKL2}]:
\begin{equation}
\begin{alignedat}{3}
    \frac{\partial \mathbf{a}_2}{\partial b_{2+}}&=\frac{k_\perp}{2}\mathbf{a}_2(m-1),&\qquad \frac{\partial \mathbf{a}_2}{\partial b_{2-}}&=-\frac{k_\perp}{2} \mathbf{a}_2(m+1),&\qquad \frac{\partial \mathbf{a}_2}{\partial b_{23}}&=ik_3 \mathbf{a}_2,\\
    \frac{\partial \mathbf{a}^*_1}{\partial b_{1+}}&=-\frac{k_\perp}{2}\mathbf{a}^*_1(m+1),&\qquad \frac{\partial \mathbf{a}^*_1}{\partial b_{1-}}&=\frac{k_\perp}{2}\mathbf{a}^*_1(m-1),&\qquad \frac{\partial \mathbf{a}^*_1}{\partial b_{13}}&=-ik_3 \mathbf{a}^*_1.
\end{alignedat}
\end{equation}
Applying these relations to the expression in the square brackets in \eqref{AA_BB}, we arrive at rather bulky formula
\begin{equation}\label{exact_int}
\begin{split}
    &(P_{01}+P'_{01})(P_{02}+P'_{02})(\mathbf{a}_1^*\dot{\spx}_1) (\mathbf{a}_2\dot{\spx}_2)+\\
    &+k_0^2\big[(\spa_1^*\spa_2) (\dot{\spx}_1-n_3\spe_3,\dot{\spx}_2-n_3\spe_3) -(\mathbf{a}_1^*,\dot{\spx}_2-n_3\spe_3)(\spa_2,\dot{\spx}_1-n_3\spe_3)\big]+\\
    &+\frac{ik_0k_\perp}{2} \big[ (\spa_1^*(m-1),(\dot{\spx}_2-n_3\spe_3)a_{2-}-\dot{x}_{2-}\spa_2) -(\spa_1^*(m+1),(\dot{\spx}_2-n_3\spe_3)a_{2+}-\dot{x}_{2+}\spa_2) -\\
    &-((\dot{\spx}_1-n_3\spe_3)a^*_{1-}-\dot{x}_{1+}\spa^*_1,\spa_2(m-1)) +((\dot{\spx}_1-n_3\spe_3)a^*_{1+}-\dot{x}_{1-}\spa^*_1,\spa_2(m+1))\big]+\\
    &+\frac{k^2_\perp}{4}\big[a^*_{1-}(m+1)a_{2+}(m-1) +a^*_{1+}(m-1)a_{2-}(m+1)-a^*_{1+}(m+1)a_{2+}(m+1)-\\
    &-a^*_{1-}(m-1)a_{2-}(m-1) +2(\spa_1^*(m+1),\spa_2(m+1)) +2(\spa_1^*(m-1),\spa_2(m-1))\big],
\end{split}
\end{equation}
where $n_3=k_3/k_0$. Now, take into account that, in the region where the radiation of an ultrarelativistic particle is concentrated,
\begin{equation}\label{estimates}
    k_0/\e\lesssim1,\qquad |\dot{x}_\pm|\sim \vk/\gamma,\qquad|\dot{x}_3|\approx1,\qquad|\dot{x}_3-n_3|\lesssim\vk^2/\gamma^2,\qquad |n_\perp|\lesssim \vk/\gamma,\qquad n_3\approx1,
\end{equation}
where $n_\perp=k_\perp/k_0$. It follows from the explicit expressions for the mode functions [(13), \cite{BKL2}] that
\begin{equation}
    |a^*_\pm|\sim |a_\pm|\lesssim\ga/\vk,\qquad |a_3|\sim |a^*_\pm a_\mp|\sim1.
\end{equation}
Expanding the scalar products in \eqref{exact_int} and neglecting the terms of order $\vk/\gamma\ll1$ in comparison with the main contribution, we obtain
\begin{equation}\label{app_int}
\begin{split}
    e^{ik_3(b_{23}-b_{13})}&\Big\{(P_{01}+P'_{01})(P_{02}+P'_{02})(\tfrac12[\dot{x}_{1-} a^*_- +\dot{x}_{1+} a^*_+]+\dot{x}_{13}a^*_3) (\tfrac12[\dot{x}_{2+} a_- +\dot{x}_{2-} a_+]+\dot{x}_{23}a_3)+\\
    &+\frac{k_0^2}{4}\big[(\dot{x}_{1+}a^*_+-in_\perp a^*_+(m-1)) (\dot{x}_{2-}a_+ +in_\perp a_+(m-1))+ \\
    &+(\dot{x}_{1-} a^*_-+in_\perp a^*_-(m+1)) (\dot{x}_{2+}a_--in_\perp a_-(m+1)) \big]\Big\}.
\end{split}
\end{equation}
Setting $q_{1,2}=q_i$ and taking into account the common factor in \eqref{AA_BB}, we deduce \eqref{prob_by_Dir}.



\end{document}